\pdfoutput=1
\documentclass[fleqn,usenatbib]{mnras}

\usepackage{newtxtext,newtxmath}

\usepackage[T1]{fontenc}

\DeclareRobustCommand{\VAN}[3]{#2}
\let\VANthebibliography\thebibliography
\def\thebibliography{\DeclareRobustCommand{\VAN}[3]{##3}\VANthebibliography}

\usepackage{graphicx}	
\usepackage{subfigure} 
\usepackage{float}
\usepackage{amsmath}	
\usepackage{xcolor}
\usepackage{tabularx}

\title[LF turnover on 21-cm signal]{Impact of the turnover in the high-z galaxy luminosity function on the 21-cm signal during Cosmic Dawn and Epoch of Reionization}
\author[Zekang Zhang et al.]{Zekang Zhang$^{1}$\thanks{E-mail: zhangzekang@shao.ac.cn}, 
Huanyuan Shan$^{1,2}$\thanks{E-mail: hyshan@shao.ac.cn}, 
Junhua Gu$^{3}$,
Qian Zheng$^{1}$,
Yidong Xu$^{3}$, 
Bin Yue$^{3}$, \newauthor
Yuchen Liu$^{1}$, 
Zhenghao Zhu$^{1}$,
Quan Guo$^{1}$\\
$^{1}$Shanghai Astronomical Observatory, CAS, 80 Nandan Road, Shanghai, China\\
$^{2}$University of Chinese Academy of Sciences, Beijing 100049, China \\ 
$^{3}$National Astronomical Observatory, Chinese Academy of Sciences, Beijing 100101, China}

\date{Accepted XXX. Received YYY; in original form ZZZ}

\pubyear{2015}

\begin{document}
\label{firstpage}
\pagerange{\pageref{firstpage}--\pageref{lastpage}}
\maketitle

\begin{abstract}
The shape of the faint-end of the high-z galaxy luminosity function (LF) informs early star formation and reionization physics during the Cosmic Dawn and Epoch of Reionization. Until recently, based on the strong gravitational lensing cluster deep surveys, the Hubble Frontier Fields (HFF) has found a potential turnover in the ultraviolet (UV) LF at z$\sim$6. In this paper, we analyze the contribution of extremely faint galaxies with the magnitude larger than the turnover magnitude in LF to cosmic reionization. We apply the measurement from HFF to our suppressed star formation efficiency model, including three free parameters: halo mass threshold $M_t$, curvature parameter $\beta$ and a UV conversion factor $l_{\rm UV}$. According to our fit of 68\% confidence level, the high-redshift star formation in haloes smaller than $ M_t=1.82^{+2.86}_{-1.08}\times10^{10} \rm M_{\odot}$ is found to be dampened. The turnover magnitude $\rm \gtrsim -13.99-2.45$, correspondingly the halo mass $\lesssim(4.57+20.03)\times10^{9} \rm M_{\odot}$. We find that the absorption trough in the global 21-cm signal is sensitive to our SFE model parameters. Together with ($\beta$, $l_{\rm UV}$) = ($2.17^{+2.42}_{-1.72}$, $9.33^{+0.43}_{-0.42} \rm ~erg~yr ~s^{-1}M_{\odot}^{-1})$, the trough locates at $\sim$ $134^{+10}_{-17}$ $\rm MHz$ with an amplitude of $\sim$ $-237^{-6}_{+7}$ $\rm mK$, compared to (106\rm MHz, -212\rm mK) in the absence of turnover. Besides, we find that the star formation of faint galaxies has also an impact on the 21-cm power spectra. The best fitting peak power decreases by $\sim4\%$ and shifts towards smaller scales from $0.88 h \rm Mpc^{-1}$ to $0.91 h \rm Mpc^{-1}$. According to our calculation, such impact is distinguishable with the forthcoming Square Kilometre Array.
\end{abstract}




\section{Introduction}

In recent years, studies for early galaxies grant insights into the astrophysics of the first generation of sources before the epoch of reionization (EoR). The main probe for galaxy studies in the early universe is the rest-frame ultraviolet (UV) luminosity function (LF), which is the volume number density per unit luminosity of all galaxies but the dust-obscured ones (e.g., \citealt{wang1996internal}; \citealt{adelberger2000multiwavelength}). By comparing it with the halo mass function, we can learn about the efficiency of star formation as a function of halo mass and redshift, which therefore provides information about gas cooling mechanism and the active galactic nuclei (AGN) and supernovae feedback processes (e.g., \citealt{van2003linking,vale2004linking,moster2010constraints,behroozi2013average,birrer2014simple}). Assembling the UV LF at high redshift is especially important to study the impact of high-z galaxies on the reionization history of the Universe, since early star-forming galaxies are seen as the major sources \citep{jiang2020quasar} to drive reionization of the intergalactic medium (IGM). 

The main progress studying high-z galaxies was yet made by deep surveys with Hubble Space Telescope (HST), currently reaching a limit around an absolute magnitude of $M_{\rm UV}$ $\sim$ $-17$, with healthy samples up to redshift $z\sim8$ \citep{bouwens2015uv,finkelstein2015evolution}. With the aid of massive galaxy clusters, strong gravitational lensing effect can magnify even fainter background galaxies. Now the Hubble Frontier Fields (HFF) programs are able to boost HST's capabilities to reach the deepest observations down to $\sim$ $29$ limit in optical and near-infrared (NIR) bands. Based on the images from HST, several groups constructed cluster mass models to interpret high-z observations, e.g., the galaxy UV LF, where at the very faint end potential turnover feature has been found by \citep{bouwens2017z, ishigaki2018full, atek2018extreme, bouwens2022}, which could be due to star formation inefficiency in small dark matter halos. This differs from previous studies which reveal a steep faint-end slope \citep{bunker2010contribution, oesch2010evolution,bouwens2011ultraviolet,mclure2013new}.

As though far more crude than these direct observations, IGM-based constraints could be a supporting approach to track high-z galaxies' evolution history. Observations of the Gunn-Peterson trough in the specta of high-z quasars suggest that the IGM is highly ionized by $z\simeq6$ (e.g., \citealt{fan2006}, \citealt{bouwens2015reionization}). Besides, the Thomson Scattering optical
depth between the observer and the cosmic microwave background
(CMB) is obtained by integrating free electron fraction along the line of sight. Using the simple tanh model for reionization, the $Planck$ data \citep{collaboration2020planck} prefer a late and fast phase transition from neutral to ionized.

Besides, redshifted 21-cm emission from neutral hydrogen hyper transition is expected to further push the study of both high-z galaxies and reionization (e.g., \citealt{madau199721,furlanetto2006global}). The sky-averaged 21-cm signal \citep{shaver1999can} offers independent constraints on the period of cosmic dawn (CD) and EoR, now being targeted by several ground-based experiments like BIGHORNS \citep{sokolowski2015}, SCI-HI \citep{voytek2014probing}, SARAS \citep{subrahmanyan2021saras} and LEDA \citep{price2018design}, as well as lunar-orbiting experiments like DAPPER \citep{burns2019dark} and DSL \citep{chen2021discovering}. It traces the volume-averaged ionization history and spin temperature history of the neutral hydrogen. The first detection of the global signal is recently reported from the Experiment to Detect the Global Epoch of Reionization Signature (EDGES; \citealt{bowman2018absorption}), whose result implies much colder IGM (e.g., \citealt{2018Natur.555...71B}) or excess background radiation \citep{2018ApJ...858L..17F} during CD, however, is inconsistent with prediction from standard cold dark matter cosmology \citep{xu2021maximum}. SARAS \citep{singh2021detection} recently claims that the best-fit profile of EDGES is rejected with 95.3\% confidence and is not evidence for new astrophysics or non-standard cosmology. Another exciting possibility is 21-cm tomography of the high-redshift IGM, in which one can map the large-scale distribution of neutral hydrogen. However, such signals are sufficiently weak that direct distribution map of HI regions is difficult to probe. Statistical measurements such as 21-cm power spectrum, being targeted by low frequency interferometers like the 21 CentiMeter Array (21CMA; \citealt{zheng2016radio}), the Giant Metrewave Radio Telescope (GMRT; \citealt{paciga2013simulation}), the Precision Array for Probing the Epoch of Reionization (PAPER; \citealt{parsons2014new}), the Murchison Widefield Array (MWA; \citealt{dillon2014overcoming}), LOw Frequency Array (LOFAR; \citealt{patil2017upper}), the Hydrogen Epoch of Reionization Array (HERA; \citealt{deboer2017hydrogen}), and the upcoming experiment Square Kilometer Array (SKA; \citealt{koopmans2015cosmic}), trace density variations and the evolution of $\rm H_{\rm II}$ regions, and can thus constrain the properties of sources. 

The combination of high-z galaxy and 21-cm signal can provide more and complementary information of cosmic reionization. \citet{mirocha2017global} build a model for the global 21-cm signal in the context of the high-z galaxy LF, and extended it in order to explain the results of EDGES \citep{mirocha2019does}. {\citet{park2019astrophysics}} discussed the constraints on the astrophysics of the first galaxies with both high-z and 21-cm observations. They find that the UV galaxy properties can be constrained at the level of $\sim 10\%$ or better if the LF turns over at an absolute magnitude brighter than $M_{\rm UV}<$-13.

 In this paper, we constrain the turnover in the faint-end of the current stellar population's UV LF with HFF observations and further investigate the impact of the turnover in LF on the redshifted 21-cm emission from neutral hydrogen, including both the global 21-cm signal and power spectrum at $z\sim 6$. This paper is organized as follows. We first briefly describe the HFF observation in Sec. \ref{Observation}. In Sec. \ref{Methods} we review the theoretical model for galaxy luminosity function and global 21-cm signal as well as 21-cm power spectrum, where we give our modification for star formation efficiency. We then give our main results in Sec. \ref{Results} and show how the observed LF turnover can affect the predictions on the EoR 21-cm observation. We conclude in Sec. \ref{Conclusions}. We use cosmological parameters from \citet{collaboration2020planck} throughout. We express all magnitudes in the AB system \citep{oke&gunnAB}.

\section{Observation}
\label{Observation}

The high-redshift UV LF has been observed by the Hubble Space Telescope over a decades-long endeavour. This has resulted in two main data catalogs dubbed the Hubble Legacy Fields and the Hubble Frontier Fields. The former consists of several deep-field surveys and has robustly probed the UV LF at the bright end, while the latter consists of observations of six cluster lenses, where faint background galaxies are magnified enough to become observable. The HFF can reach fainter objects, as those are strongly magnified by the cluster lenses, whereas lensing can introduce important uncertainties. 

HFF imaged $6$ clusters and flanking fields in the optical bands with $3$ filters F435W, F606W, and F814W by the Advanced Camera for survey and the NIR bands with $4$ filters F105W, F125W, F140W, and F160W by the Wide Field Camera Three. HST first achieves two bands for each pair cluster/parallel field and then the positions of instruments are switched. For each pair, observations were taken in HST cycles from $21$ to $23$ between $2013$ and $2016$ with $140$ orbits in total \citep{atek2018extreme}.

With the lensing models \citep{lotz2017frontier} of the HFF data, the faintest high-redshift galaxies can be discovered. Furthermore, it can help us to measure the high-z UV LF \citep{atek2014probing,zheng2014young,yue2014ultra,ishigaki2015hubble,kawamata2016precise,castellano2016first,laporte2016young,ishigaki2018full,bouwens2017z,livermore2017directly}. Current detection limits can be down to $M_{\rm UV}$ $\sim$ -13 and even to $M_{\rm UV}$ $\sim$ -12 \citep{livermore2017directly}. The UV LF from different groups are in good agreement on the bight end with $M_{\rm UV} < $ -17, however, significant discrepancies can be found at the faint end, where only sources at high magnification region can be resolved. \cite{bouwens2017z} find potential turnover in the LF at $M_{\rm UV}$ > -15 and a faint-end slope of $\alpha = -1.91 \pm 0.04$. While, \citet{livermore2017directly} shows considerably higher values of the LF at $M_{\rm UV} >$ -17 and \cite{ishigaki2018full} find a steeper faint-end slope. \cite{livermore2017directly} also suggest a steep faint-end slope of $\alpha$ $\sim$ $2.10\pm0.03$ and a strong proof against a potential turnover at $M_{\rm UV} <$ -12.5.

With end-to-end simulations that account for all lensing effects and systematic uncertainties by comparing several mass models, \cite{atek2018extreme} found that tight constraints on the LF fainter than $M_{\rm UV}$ $\sim$ -15 remain impossible, as the $95\%$ confidence interval (CL) indicates a turnover although a steep faint-end slope is also permitted. Such a turnover in the faint-end of the UV LF could result from the inefficiency of star formation in small dark matter halos (e.g., \citealt{jaacks2013impact,gnedin2016cosmic,yue2016faint}), which can make an impact on the cosmic reionization history.

\section{Methods}
\label{Methods}

In this section, we briefly outline the model that connects high-z galaxy population with redshifted 21-cm emission (Mirocha et al. 2016), including three parts: (i) constructing the UV LF of high-z galaxy by abundance matching as well as the relation between dark matter halo mass and galaxy luminosity; (ii) generating the global 21-cm signal with the galaxy LF. With this model, we examine the impact of the observed turnover in the faint-end of the UV LF on the global 21-cm signal. (iii) calculating the 21-cm power spectrum. Assuming the UV LF model of galaxy are connected to the host halos, we modify the halo mass function (HMF) to generate the turn-over LF. Since the global 21-cm signal traces the volume-averaged ionization and thermal histories in time, and thus constraints on the whole reionization history. The extrapolation of the current HFF observation at z $\sim$ 6 is incomplete. Therefore we further calculate the power spectrum to check the influence of the turn-over LF.

\begin{figure*}
    \begin{subfigure}
        \centering
        \includegraphics[width=0.45\textwidth]{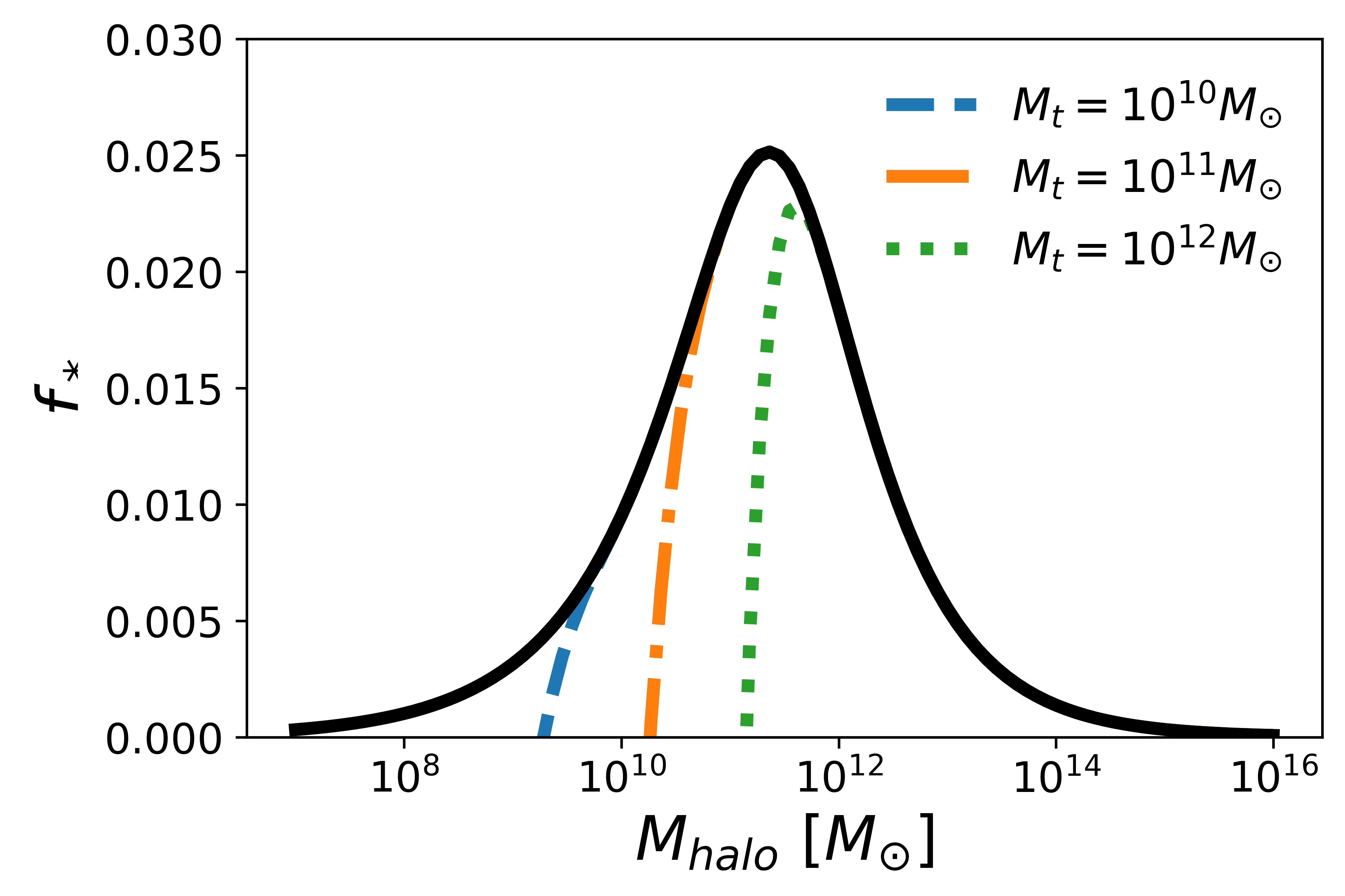}
        \includegraphics[width=0.45\textwidth]{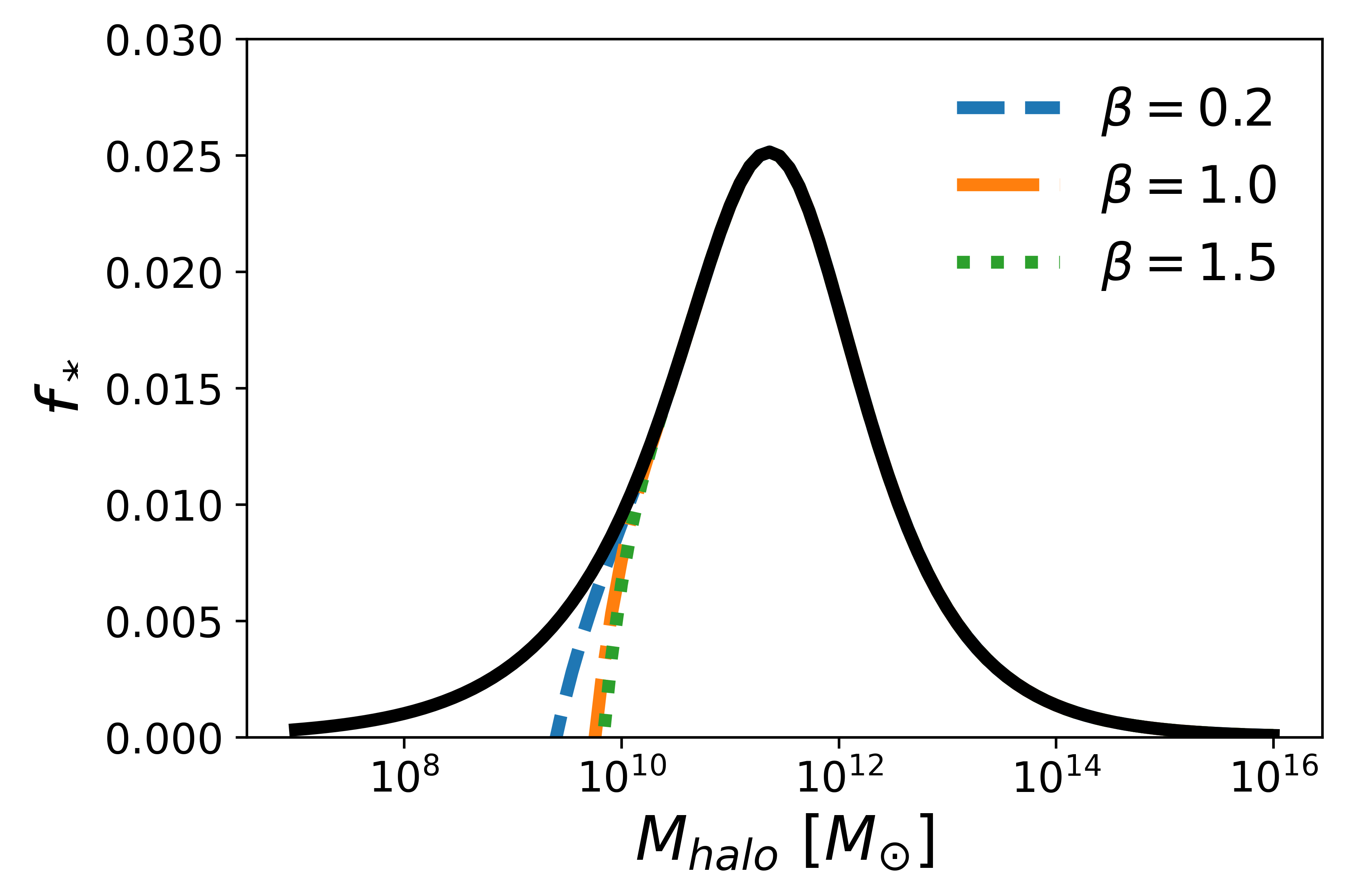}
    \end{subfigure}
    \begin{subfigure}
        \centering
        \includegraphics[width=0.44\textwidth]{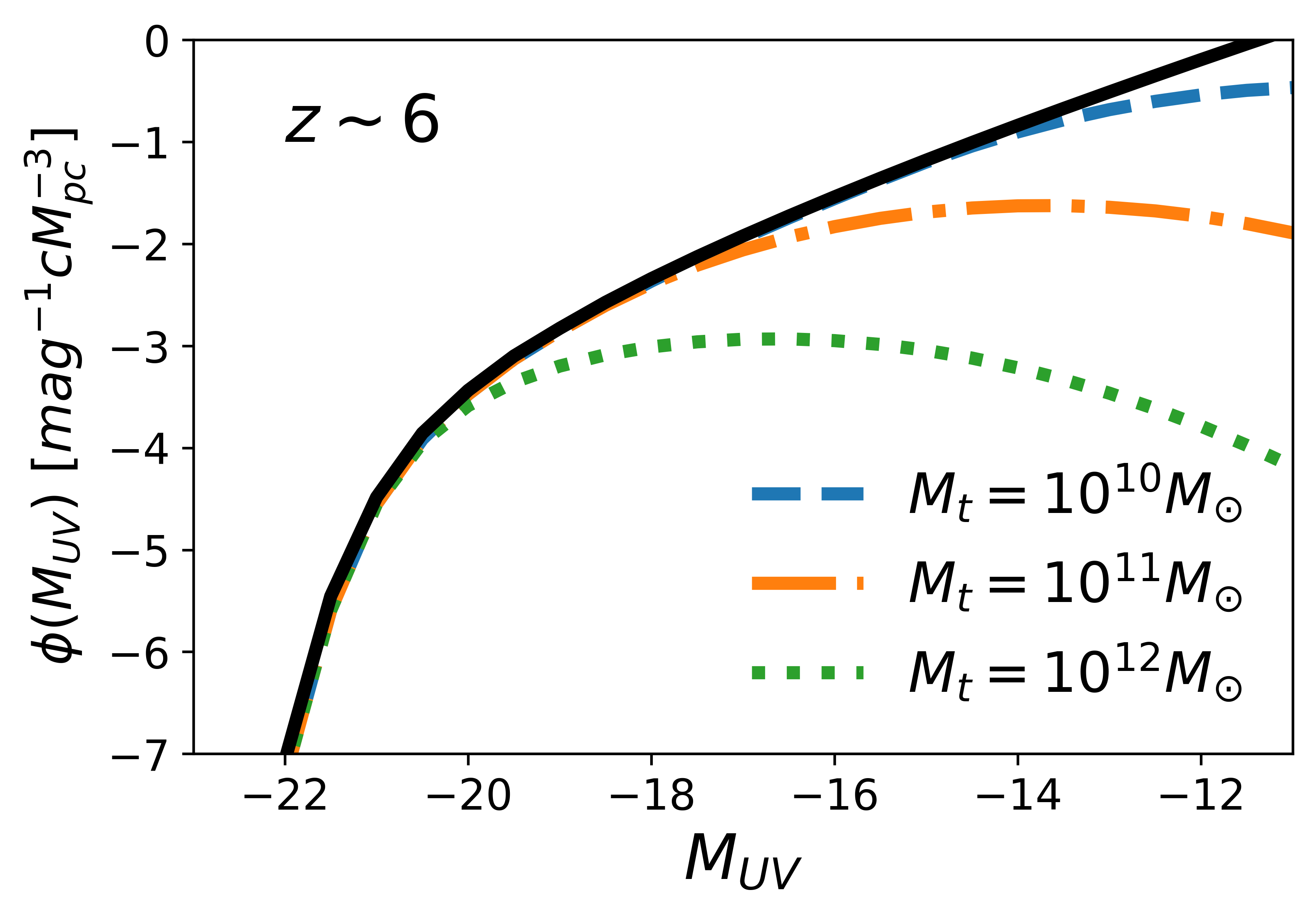}
        \includegraphics[width=0.44\textwidth]{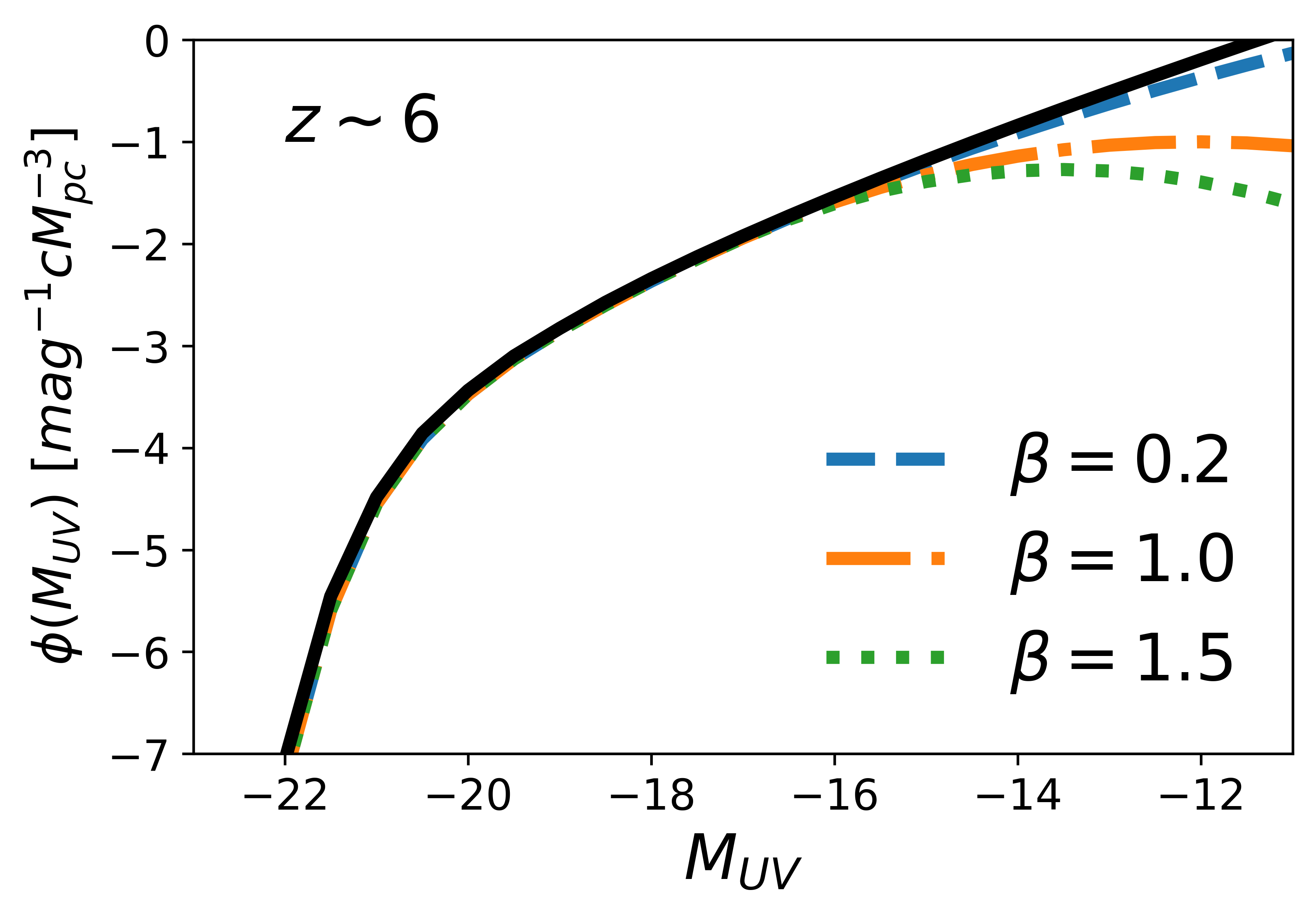}
    \end{subfigure}
    \begin{subfigure}
        \centering
        \includegraphics[width=0.45\textwidth]{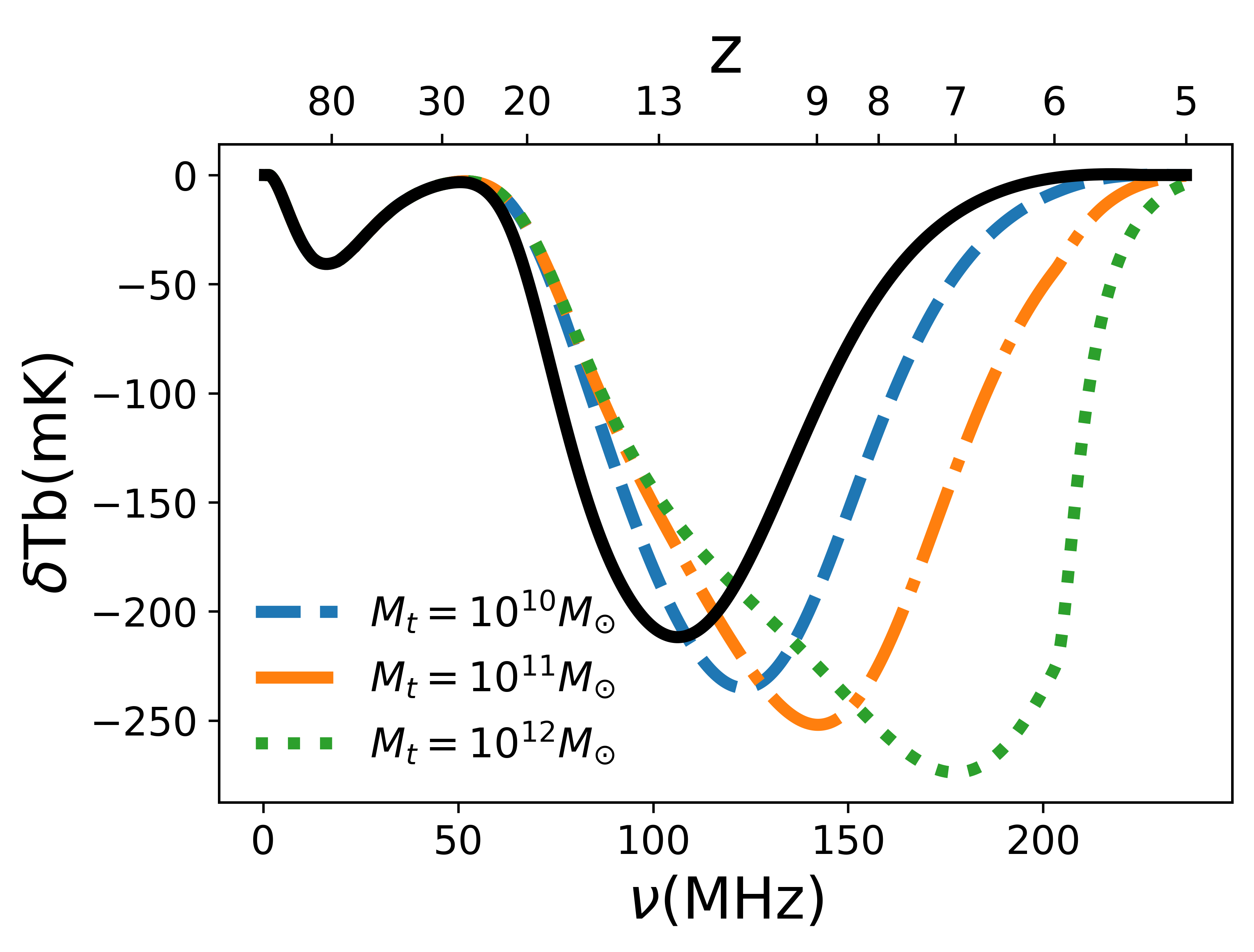}
        \includegraphics[width=0.45\textwidth]{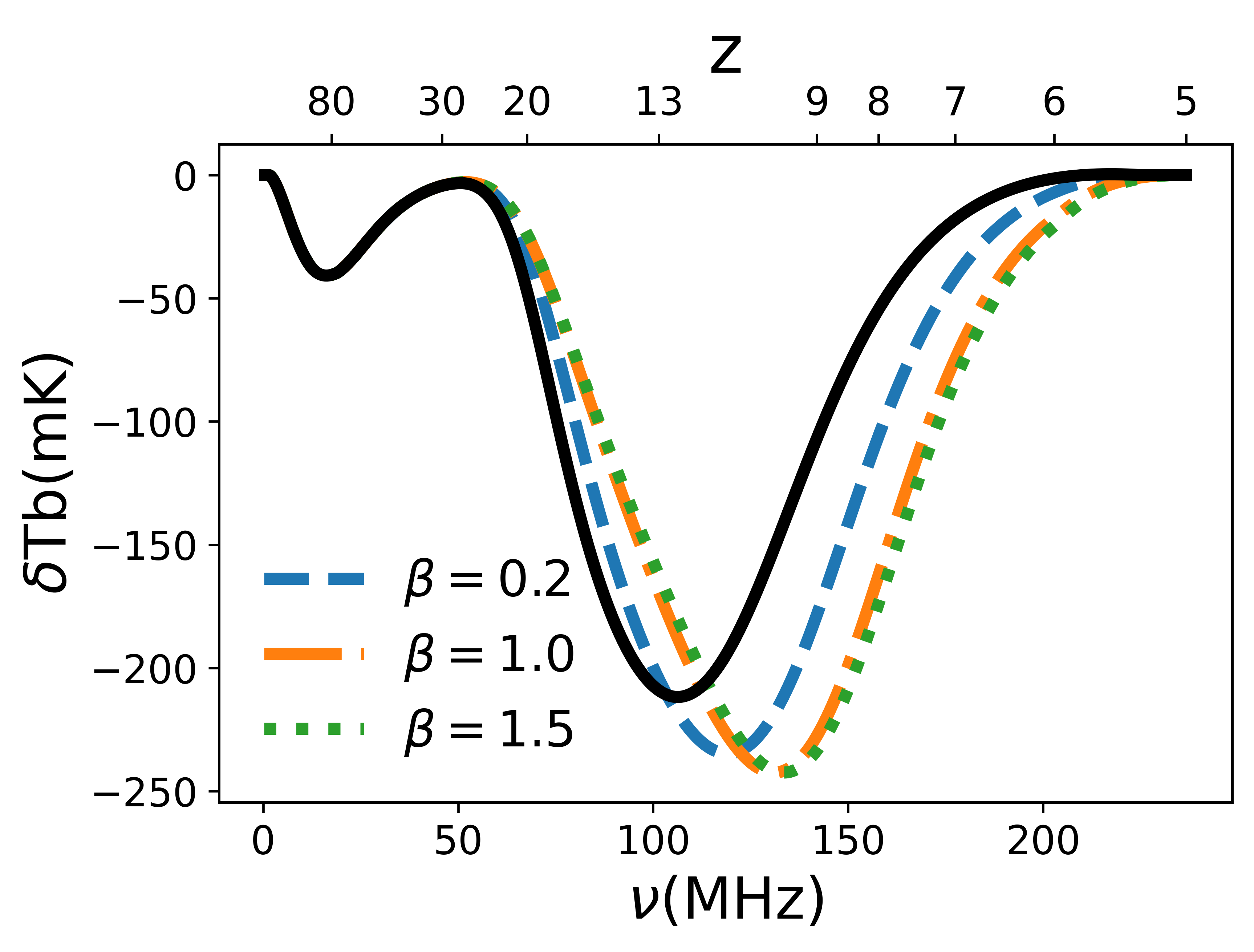}
    \end{subfigure}
    \begin{subfigure}
        \centering
        \includegraphics[width=0.455\textwidth]{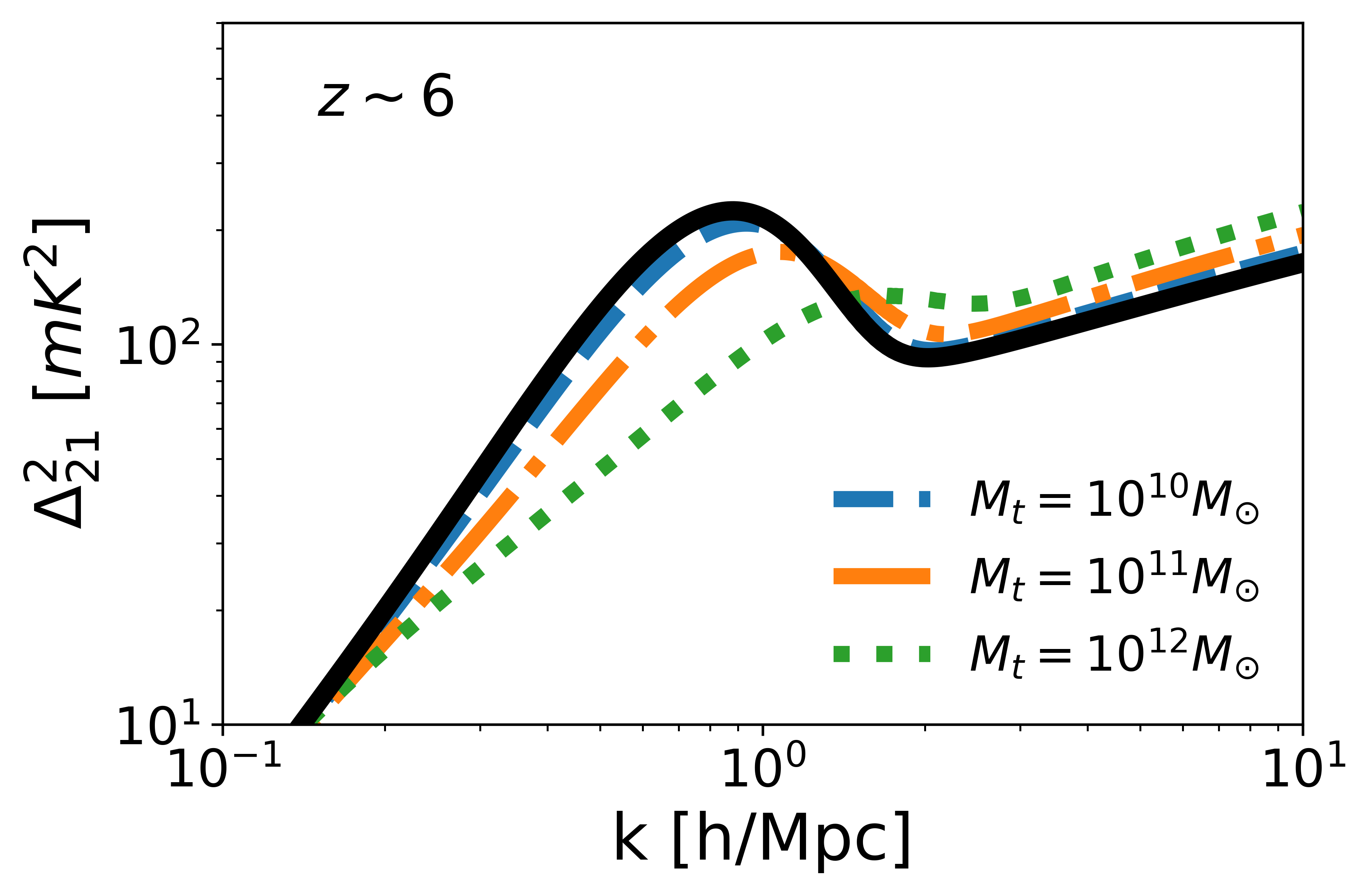}
        \includegraphics[width=0.455\textwidth]{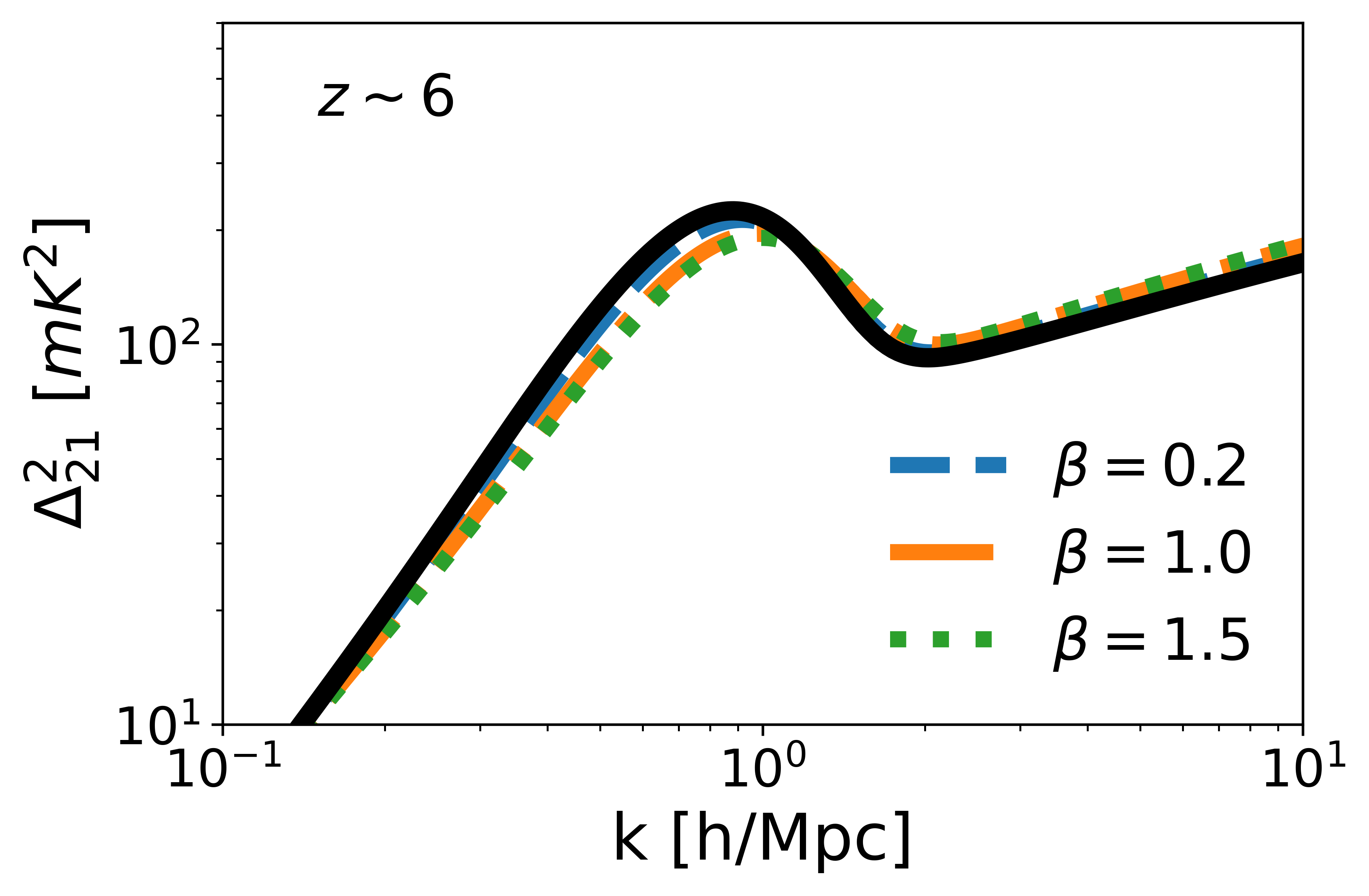}
    \end{subfigure}
    \caption{Response of the global 21-cm signal and power spectrum to changes of two free parameters $M_t$ (left column) and $\beta$ (right column). Solid thick curves are our fiducial model. The left column describes changes when we set $M_t$ as $1\times10^{10}$, $1\times10^{11}$ and $1\times10^{12}$ $\rm M_{\odot}$ ($\beta$ fixed at $1.0$) while the right column is when we change $\beta$ as 0.2, 1.0 and 1.5 ($M_t$ fixed at $3\times10^{10} \rm M_{\odot}$).}
    \label{fig:1}
\end{figure*}

\subsection{UV LF}
\label{uvlf}

The property of galaxies has tight connections with their host dark matter halos. More luminous and massive galaxies tend to reside in more massive halos. In abundance matching method, the star formation in galaxies is assumed to be a monotonic function of the host halo's mass, which is consistent with the observed trend that the galaxies' clustering strengthens with their UV luminosity, similar to that of the clustering strength of halos increasing with mass. The number of galaxies is assumed to equal the number of host halos, where the subhalo population is neglected since the impact of such substructure within a halo is found to be negligible \citep{mason2015galaxy,mashian2016empirical}. Therefore, the galaxy luminosity can be constructed as a function of halo mass and redshift,
\begin{equation}
d\phi(L_h) = \frac{dn(M_h, z)}{dM_h} \left ( \frac{dL_h}{dM_h} \right )^{-1} dL_h, 
\end{equation}
where $\phi$ is the galaxy LF, $n(M_h, z)$ is the halo number density and  $dn(M_h,z)/dM_h$ is the mass function of dark matter halos. Here we adopt the Sheth-Tormen mass function in the calculation \citep{sheth1999large}.

Assuming young and massive stars in high-z galaxies dominate the UV production, the intrinsic luminosity of galaxies can then be modelled through the star formation rate with a conversion factor $l_{\nu}$ (luminosity per unit star formation rate) \citep{kennicutt1998,madau2014},
\begin{equation}
L_{h, \nu} = \dot{M}_*(M_h, z) l_\nu, 
\end{equation}
the star formation is fuelled by inflow of IGM gas into galaxies \citep{sun2016constraints},
\begin{equation}
\dot{M}_*(M_h, z) = f_*(M_h, z)\dot{M}_b(M_h, z), 
\end{equation}
in which the baryonic mass accretion rate can be well approximated as $\dot{M}_b \propto M_h(1+z)^{5/2}$ \citep{mcbride2009mass,dekel2013toy}.

The analysis from both simulations and observations (e.g.,  \citealt{wechsler2018connection,sun2016constraints,behroozi2019universemachine,moster2018emerge}) indicates that the star formation in both high-mass and low-mass halos would be less efficient as a result of baryonic feedback, such as active galactic nuclei ejection and supernovae shocks. Therefore, a double power-law (DPL) model of star formation efficiency (SFE) is assumed,
\begin{equation}
    f_{*,\rm dpl}(M_h)=\frac{f_{*,0}}{\left (\frac{M_h}{M_p} \right )^{\gamma_{lo}}+\left (\frac{M_h}{M_p} \right )^{\gamma_{hi}}},
\end{equation}
where $f_{\star,0}\geq0$ describes its amplitude, $\gamma_{lo}\leq0$ and $\gamma_{hi}\geq0$ describe the slope of the low and high mass end of the UV LF and $M_p$ sets the halo mass where SFE peaks. According to \citep{mirocha2017global}, the four parameters in the model are taken to be ($\emph{f}_{*,0}$, $M_p$, $\gamma_{lo}$, $\gamma_{hi}$) = ($0.05$, $2.8\times10^{11} \rm M_{\odot}$, $0.49$, $-0.61$), which is derived from the LF measurements at $z\sim5,7~\rm and~8$ \citep{bouwens2015uv}.

This redshift-independent star formation model is consistent with current observations at the range of $0\lesssim z \lesssim8$. However, based on this framework, the new finding at the faintest limit from HFF can not be recovered. The observed turnover is believed to result from SN and radiative feedback during reionization, which has a significant impact on the surrounding environment and decreases the necessary gas supply for star formation. Based on the DPL model, we reconstruct the SFE at the low-mass end where $M_h<M_t$ as, 
\begin{equation}
    f_*(M_h)=\frac{1}{M_h}\left [ C+\int_{M_{h,\rm min}}^{M_h} \frac{1}{T}(f_{*,\rm dpl}+M_hf_{*,\rm dpl})dM_h \right ],
    \label{con:sfe}
\end{equation}
where $\rm C$ is the normalization, and $f_{*,\rm dpl}$ refers the original double power law form. The virial temperature threshold is fixed at $10^4 \rm K$. This sets the lower bound of the integral as the corresponding atomic cooling halo mass $M_{h,\rm min}$, which is supposed to be redshift-dependent \citep{barkana2001}. This formula induces a steep declination in the SFE at the low-mass end, and shows on LF as a multiplicative term T where the galaxies reside in halos smaller than $M_t$,
\begin{equation}
    T = \exp{[-\beta \log_{10}(M_h/M_t)^2]},
    \label{con:turnover}
\end{equation}
in which $\beta$ is the curvature parameter. When $\beta>0$, the UV LF have a downward turnover and an upward turnover for $\beta<0$. In our fiducial (baseline) scenario, $\beta=0$, which means no turnover is introduced.

Dust extinction reduces the amplitude of UV LFs brighter than $\rm M_{UV}\sim-20$ at $z\gtrsim6$ (e.g., \citealt{yung2019dust,vogelsberger2020dust}). This could induce bias on the calibration of UV LFs at the bright end. However, bright galaxies are rare at high redshifts and contribute little to the emissivity. Therefore, dust extinction is not considered in the model by \cite{mirocha2017global} as it has only a minor impact on the global 21-cm signal of $\sim0.1 \%$ according to our estimation. Our SFE model acts as an integration of DPL between $M_{h,\rm min}$ and $M_h$. When $M_h>M_t$, $T=1$ and Eq. \eqref{con:sfe} equals to DPL. Since the turnover is expected to happen much fainter than $\rm M_{UV}\sim-20$, the integration will not introduce further biased results and we ignore dust in this work as well.

In order to model the global 21-cm signal, we have to extrapolate the current results to lower mass halos and higher redshifts, during which the feedback mechanisms that suppress small galaxies' star formation are more complicated and remain unclear. In this work, we only focus on how the observations by HFF at z $\sim$ 6 would affect the reionization histories and ignore the redshift evolution. Besides, the power spectrum of 21-cm brightness temperature at z $\sim$ 6 is also considered, which is discussed in detail in Sec. \ref{21cmps}.

\subsection{Global 21-cm signal}

We employ the two-zone model in which the IGM is partitioned into a fully ionized phase and the 'bulk' IGM. The global 21-cm signal is then the volume-averaged brightness temperature of the bulk IGM (e.g., \citealt{furlanetto2006global,pritchard2010constraining}),
\begin{equation}
    \delta T_b \simeq 27(1-\Bar{x_i}) \left (\frac{\Omega_{b,0}h^2}{0.023} \right ) \left (\frac{0.15}{\Omega_{m,0}h^2} 
    \frac{1+z}{10} \right )^{1/2} \left (1-\frac{T_\gamma}{T_S} \right )
\label{AvedTb}
\end{equation}
where $\Bar{x_i}=Q_{\rm H_{II}}+(1-Q_{\rm H_{II}})\emph{x}_e$ is the volume-averaged ionized fraction, $\emph{x}_e$ is the electron fraction in partially-ionized regions, $Q_{\rm H_{II}}$ is the volume-filling factor of fully-ionized regions, $T_\gamma$ is the cosmic background radiation and $T_S$ denotes the spin temperature of neutral hydrogen,
\begin{equation}
    T_S^{-1} \approx \frac{T_R^{-1}+x_c T_K^{-1}+x_{\alpha} T_{\alpha}^{-1}}{1+x_c+x_{\alpha}}.
\end{equation}

The signal depends on the time evolution of the IGM's thermal and ionization conditions. This requires the knowledge of the mean radiation background intensity $J_{\nu}$ pervading the IGM. With the luminosity function described in Sec. \ref{uvlf}, the volume-averaged emissivity can be computed via integration of the galaxy LF,
\begin{equation}
    \epsilon_{\nu}(z)=\int_{L_{\rm min}} f_{\rm esc,\nu}L_{h,\nu}\frac{d\phi(L_{h,\nu})}{dL_{h,\nu}}dL_{h,\nu},
\end{equation}
with escape fraction $f_{\rm esc,\nu}$ in relevant bands. The lower bound of the integral corresponds to $M_{h,min}$. Photons in ultraviolet and X-ray band are considered in calculation. The energy bands include $\rm Ly\alpha$ and Lyman-Werner photons at 10.2-13.6eV, Lyman continuum (LyC) photons at 13.6-24.6eV and X-ray photons at a range of 200-30000eV. In general, $l_{\nu}$ can depend on $M_h$ and z, though here we take it as a constant. Therefore the spectral energy distribution of galaxies is mass-independent and do not evolve through time. We take $l_{\rm UV}$ as a free parameter to fit as described in Sec. \ref{fit}. For X-ray sources, \citet{mirocha2014decoding} models $l_{\rm X}$ as a multi-color disc of $10 \rm M_{\odot}$ black holes, which is representative of high-mass X-ray binaries and is believed to be the most important X-ray sources in high-z galaxies. Throughout the paper, we set the escape fraction of the $\rm Ly\alpha$ and soft UV photons at $10.2-13.6 \rm eV$ and the X-ray photons as 1.0. Constraints on the escape fraction of LyC photons varied widely with redshift \citep{robertson2021escapefraction}. We adopt the results by \cite{meyer2020scape}, in which they inferred an escape fraction of $f_{\rm esc,\rm LyC}=0.23^{+0.46}_{-0.12}$ for star-forming galaxies at $z\sim5.5-6.4$ utilising two-point correlations between galaxies and IGM transmissivity in sightlines to distant quasars.

Considering the effects of redshifting and bound-free absorption by neutral gas (e.g. $\rm H_I,He_I$), $J_{\nu}$ can be calculated by solving the cosmological radiative transfer equation using $\epsilon_{\nu}$. In soft-UV band, $J_{\nu}$ determines the evolution of the electron fraction $x_e$ in the bulk IGM and $\rm Q_{H_{II}}$. $\Bar{x_i}$ then dictates the IGM optical depth. For X-ray, $J_{\nu}$ plays a major role to heat the gas, together with other heating and cooling sources (e.g. Compton heating and Hubble cooling). With it we can solve the time evolution of $T_K$ for the bulk IGM. $J_{\alpha}$ at the $\rm Ly\alpha$ frequency determines the Wouthuysen-Field effect, which couples $T_S$ to $T_{\alpha}$ and in most circumstances $T_{\alpha}\simeq T_K$ \citep{1959ApJ...129..536F}. The coupling coefficient is calculated as $x_{\alpha}=1.81\times10^{11}(1+z)^{-1}S_{\alpha}J_{\alpha}$, where $S_{\alpha}$ is a factor of order describing the atomic scatter process. $x_c$ is the collisional coupling coefficient for H-H interactions and is computed using the tabulated values in \citet{2005ApJ...622.1356Z}. All calculation were carried out with the {\it ARES }\footnote{\url{https://github.com/mirochaj/ares}} code and given in detail in \citep{mirocha2014decoding,gu2020direct}.

\begin{figure*}
    \centering
    \begin{subfigure}
        \centering
        \includegraphics[width=0.7\linewidth]{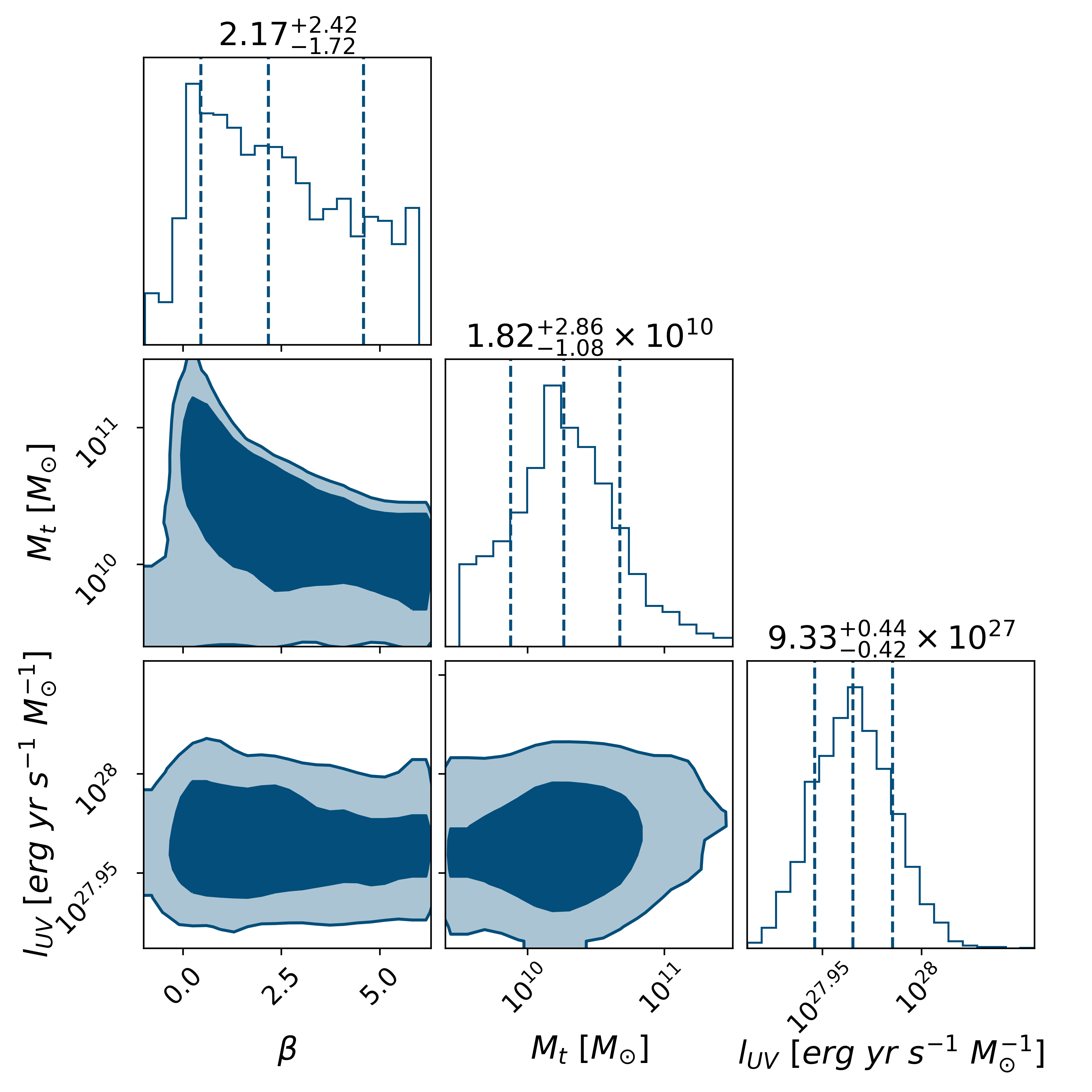}
    \end{subfigure}
    \begin{subfigure}
        \centering
        \includegraphics[width=0.48\textwidth]{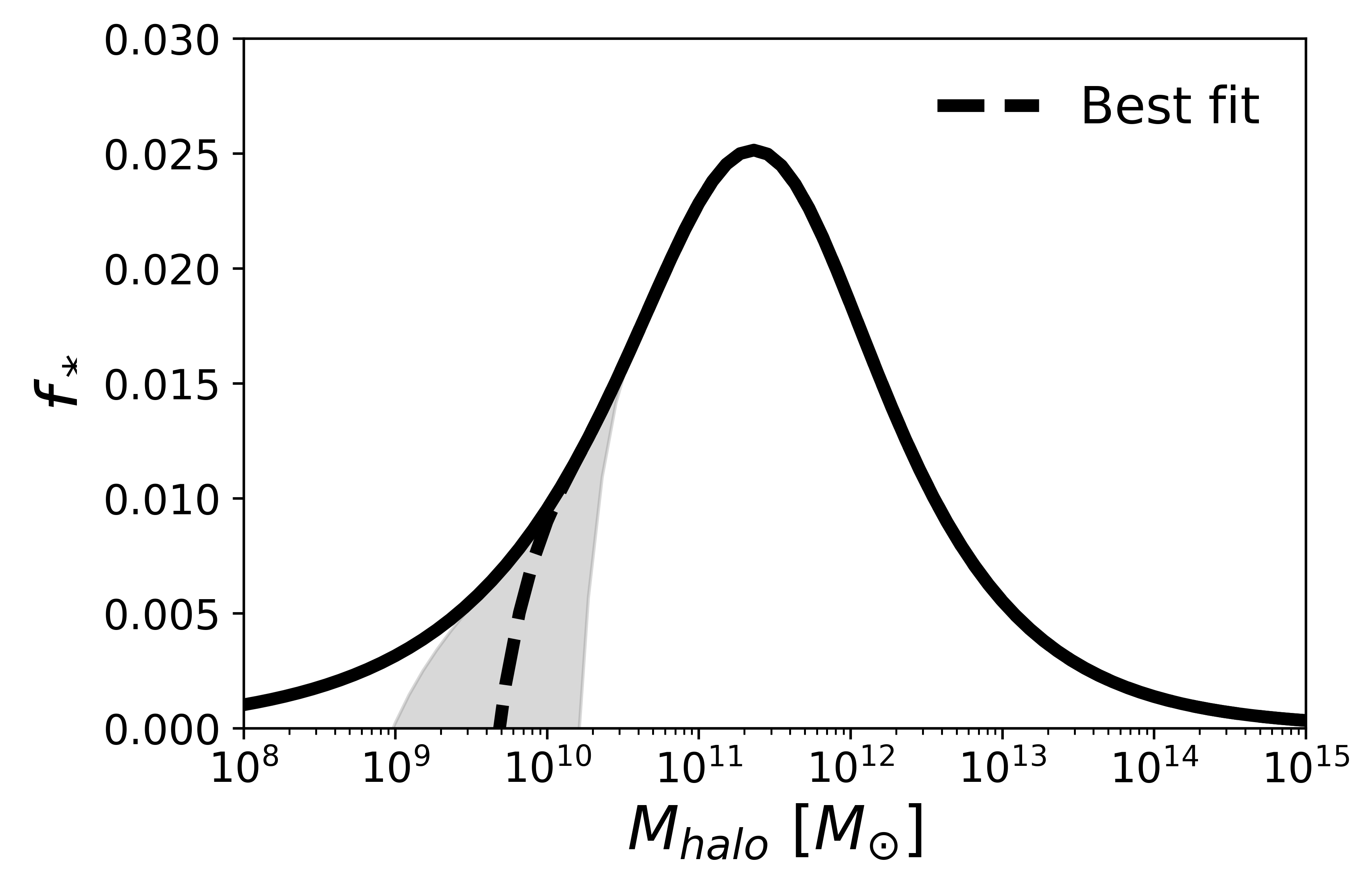}
        \includegraphics[width=0.45\textwidth]{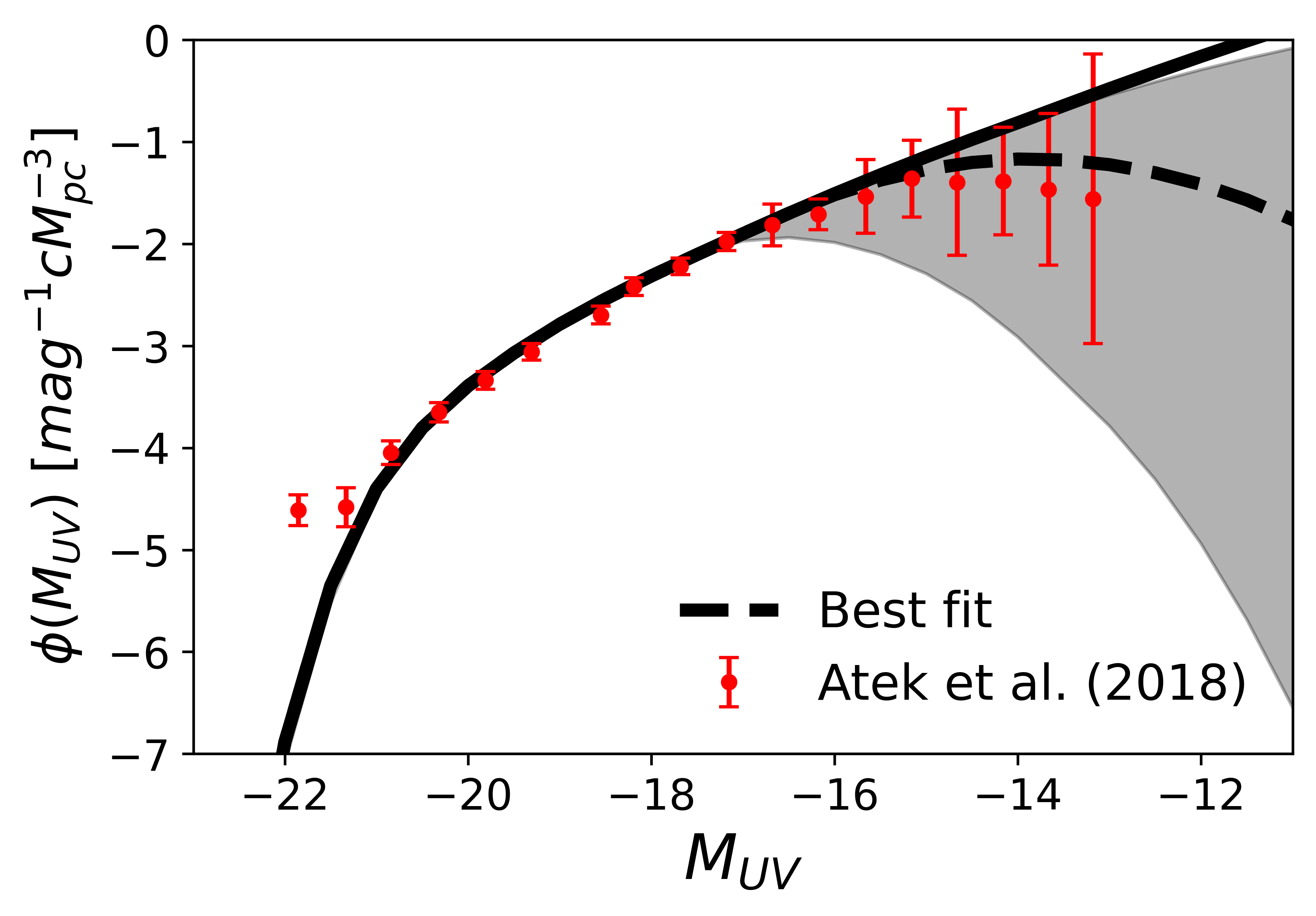}
    \end{subfigure}
    \caption{Top: Constraints on three parameters of the suppressed SFE model from the HFF deep HST imaging data (\citealt{atek2018extreme}). Contours correspond to $68$ \% and $95$ \% confidence level (CL). The boundary of $68$ \% intervals and the medians of 1D posteriors are marked as blue dashed lines, of which the exact numbers are shown as the titles.  Bottom: The corresponding results of star formation efficiency (left) and galaxy luminosity function (right). The dark solid lines represent our fiducial scenario. The black dashed lines represent when we apply the best fit parameters. The grey areas give the interval of SFE and UV LF when we take the 68\% CL boundary of parameter sets.}
    \label{fig:2}
\end{figure*}

\subsection{21-cm Power spectrum}
\label{21cmps}

Since the global 21-cm signal describes the whole history of the IGM during CD/EoR in the redshift range of ($5<z<30$), a direct extrapolation of our SFE model calibrated at z $\sim6$ could be incomplete. Therefore, 21-cm power spectrum at $z\sim 6$ can be a better tracer since it gives a tomographic description of the 21-cm field at fixed redshift. In this work, according to the UV LF model that connects galaxy population to the host halos, we simply modifies the high-z halo mass function to match the turnover on LF. 

We adopt a model for reionization associating $\rm H_{II}$ regions with large-scale overdensities \citep{furlanetto2004growth}. Different from the volume-averaged signal with averaged ionized fraction $\Bar{x}_i$ as is described in Eq. \eqref{AvedTb}, here $1-\Bar{x_i}$ is replaced by $\psi=x_{\rm H}(1+\delta)$, which represents the conjugation of the 21-cm field between the local overdensity $\delta$ and the local neutral fraction $x_{\rm H}$ \citep{madau199721}. 

We can then model the power spectrum of $\psi$ and therefore the 21-cm brightness temperature. The correlation function of $\psi$ is given from its component fields,
\begin{equation}
    \xi_{\psi}=\xi_{xx}(1+\xi_{\delta\delta})+\Bar{x}_{\rm H}^2\xi_{\delta\delta}+\xi_{x\delta}(2\Bar{x}_{\rm H}+\xi_{x\delta})
\end{equation}
where $\xi_{xx}$ is the correlation function of the $\rm H_{\rm II}$ regions in terms of the ionized bubble mass function above, $\xi_{\delta\delta}$ is the density correlation function in terms of the HMF, and $\xi_{x\delta}$ is the cross correlation between density and ionization fraction. The statistics of $\psi$ are determined by its power spectrum that is the Fourier transform of $\xi_\psi$ \citep{zaldarriaga200421},
\begin{equation}
    \langle\hat{\psi}({\bf k}_1)\hat{\psi}({\bf k}_2)\rangle=(2\pi)^3\delta({\bf k}_1+{\bf k}_2)P_{\psi}(k_1)
\end{equation}
where the power spectrum can also be given in a dimensional form as $\Delta^2_{21}(k)=P_{\psi}(k)k^3/(2\pi^2T^2_0)$ and $T_0=28[(1+z)/10]^{1/2} \rm mK$.

In this formalism, star formation efficiency is directly related to the ionizing efficiency $\zeta$ of sources. A galaxy of mass $m_{\rm gal}$ is assumed to ionize a mass $\zeta m_{\rm gal}$. $\zeta$ is usually considered to be $10\sim40$. Besides the SFE, this quantity is also governed by ionizing photon production and the escape fraction as well as the level of recombination. Therefore, if star formation in small dark matter halos are suppressed, lower ionizing efficiency in small sources are expected, which thus affects the mass function of $\rm H_{\rm II}$ regions. In this work, we embed the turnover (Eq.\ref{con:turnover}) of galaxy LF in halo mass function instead of modifying the ionizing efficiency $\zeta$. According to current measurements, the co-moving number density of dark matter halos is consistent with theoretical models and does not show any declination at low mass end. Therefore, our treatment here is an equivalent method, which can produce the same results as the modification of $\zeta$.

\begin{figure*}
    \includegraphics[width=0.7\textwidth]{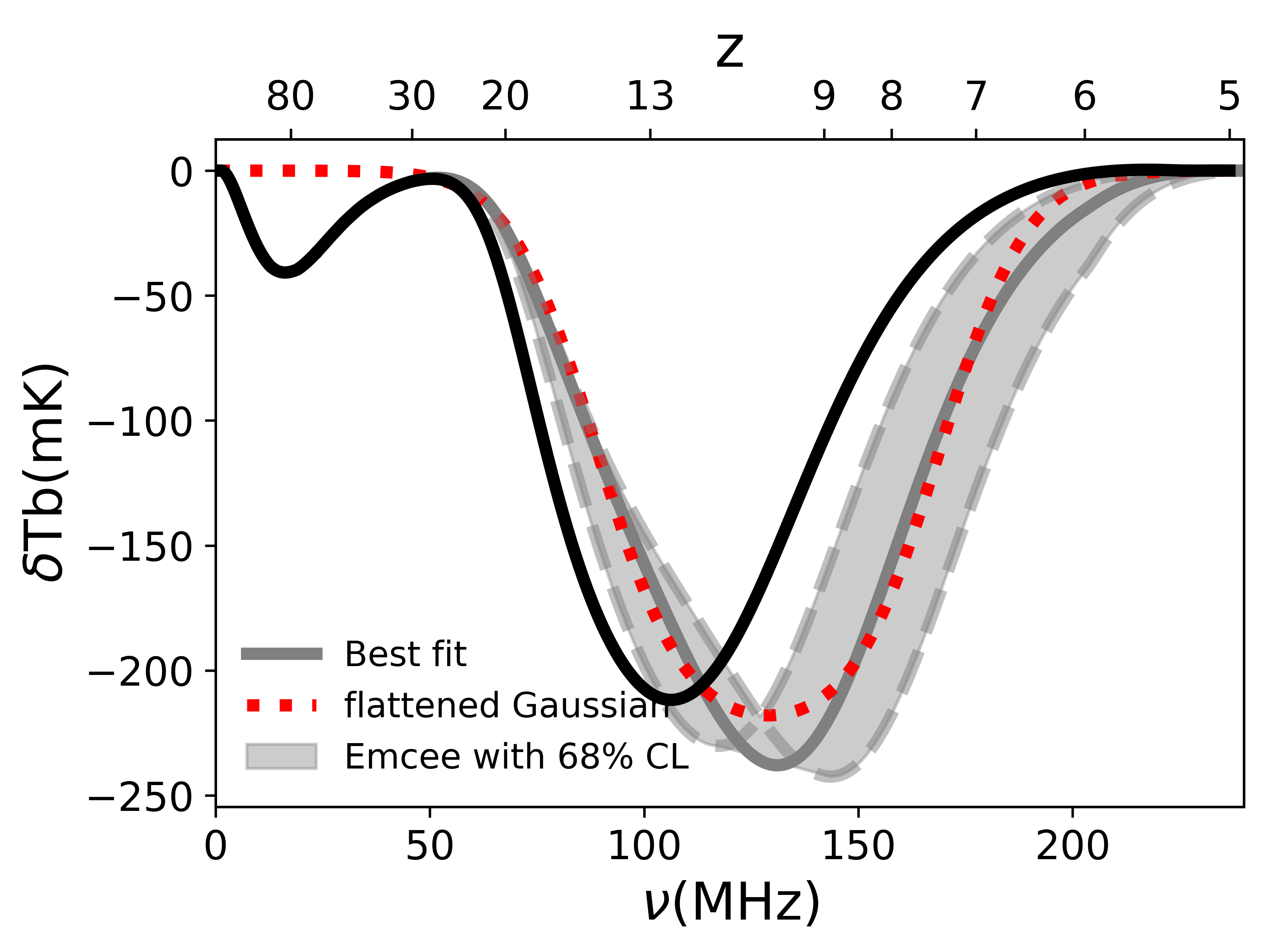}
    \caption{Models of the global 21-cm signal corresponding to the LF turnover. The solid dark curve represents our fiducial model. The red dotted curve is the fit of the solid grey curve to a flattened Gaussian profile as Eq. \eqref{eq:gauss}. The 21-cm global spectra for fiducial SFE model (dark black solid) and the model that incorporates the LF turnover derived from HFF observations (grey). Shaded area corresponds to the interval of global 21-cm signals when we take the 68\% CL boundary of parameter sets.
    }
    \label{fig:3}
\end{figure*}

\section{Results}
\label{Results}

\subsection{Impact of LF turnover on the EoR 21-cm signal}

In this section, we estimate the impact of LF turnover on the EoR 21-cm signal, including both the global 21-cm signal and 21-cm power spectrum. 

The volume-averaged emissivity, computed as a integral over galaxy UV LF, can determine the ionization and heating progress of the IGM. In the fully ionized phase, it governs the rate at which the bubbles grow. In other words, LF influences the global 21-cm signal mainly by determining the meta-galactic radiation background therefore the ionizing and thermal history of the IGM. This implies that when inducing a turnover on LF, we expect suppression in these processes. Before the X ray sources ignite, the decresing $\rm Ly\alpha$ flux will increase the spin temperature of neutral gas. After that, X ray heats the gas and $T_S$ begins to couple with the kinetic temperature of IGM. Lower X ray flux means lower $T_K$ and therefore lower $T_S$. Also the timing when X ray heating dominates will significantly delay. As a result, this will deepen the absorption trough in the global signal and shift it towards higher frequency.

Considering different star formation scenarios, we can investigate the influence of the turn-over LF model. The results are shown in Fig.~\ref{fig:1}. The effects are quite similar whether we change the mass threshold or the curvature parameter $\beta$. In general, comparing with scenario in the absence of the turnover (solid dark lines), the absorption trough in the global signal significantly deepens by $\sim 30$\rm mK with a down shift of redshift $\Delta z \sim 3$. At the mass threshold of $M_t=1\times 10^{10}, 1 \times  10^{11}, 1 \times 10^{12} \rm M_{\odot}$, the declining star formation efficiency corresponds to a shortage of faint galaxies. Increasing $M_t$ to $1 \times 10^{11} \rm M_{\odot}$ ($1 \times 10^{12} \rm M_{\odot}$) shifts the signal by $\sim 30\rm MHz$ ($\sim 70\rm MHz$) towards higher frequency and deepens it in brightness temperature by $\sim 30\rm mK$ ($\sim 80\rm mK$). For the curvature parameter $\beta=0.2,1.0,1.5$, the slope of both our SFE model and the galaxy LF's faint end gets steeper. The $\beta=1.5$ can shift the signal to higher frequency by $\sim 30\rm MHz$ while the absorption deepens by $\sim 30\rm mK$. 

In the last row of Fig.~\ref{fig:1}, we check the impact of the turnover on 21-cm power spectrum. From these panels, we can find that increasing $M_t$ to $1 \times 10^{11} \rm M_{\odot}$ slightly increases the small-scale power and moves the bubble peak to smaller scales and lower amplitude. When $M_t$ is further set as $1 \times 10^{12} \rm M_{\odot}$, the curve is flattened on all k modes. The bubble feature is nearly diminished. On the other hand, higher $\beta$ has similar but relatively minor effects on the power spectrum.

\subsection{Fitting the HFF LF measurement}
\label{fit}

\begin{table}
    \centering
    \begin{tabular}{ c c c c}
    \hline\\
    Flat prior & 68\% CI & Unit\\[2ex] 
    \hline\\
    $10^{9.5}<M_t<10^{12.5}$ & $1.82^{+2.86}_{-1.08}\times10^{10}$ & $\rm M_{\odot}$\\[2ex] 
    \hline\\
    $-1<\beta<6$ & $2.17^{+2.42}_{-1.72}$ & \\[2ex] 
    \hline\\
    $10^{27}<l_{\rm UV}<10^{29}$ & $9.33^{+0.44}_{-0.42}\times10^{27}$ & $\rm erg~yr~s^{-1} \rm M_{\odot}^{-1}$\\[2ex] 
    \hline\\
    \end{tabular}
    \caption{Priors and parameter 68 \% confidence intervals for our suppressed SFE model, including the turnover parameters $(M_t, \beta)$ and the conversion factor $l_{\rm UV}$.}
    \label{tab:fit}
\end{table}

With the measurements of galaxy LF at z $\sim$ $6$ from HFF, we can then constrain the turnover model, which can be described by 3 free parameters of $(M_t, \beta, l_{\rm UV})$. Here we take $l_{\rm UV}$ into consideration as it can be correlated with the turnover tail. For example, an improperly high $l_{\rm UV}$ can result in much higher $M_t$ and corresponding lower $\beta$, which means an earlier and flat turn at the faint end. A fixed $l_{\rm UV}$ can therefore bias the constraints on the turnover model. We fit the data using the Markov Chain Monte Carlo (MCMC) method with {\it EMCEE} \footnote{\url{https://github.com/dfm/emcee}} \citep{foreman2013emcee}, with stable results in 10000 steps of 8 chains. We adopt flat priors on each model parameter of interest, as is shown in Table \ref{tab:fit}. \cite{atek2018extreme} fixed $M_t$ (the corresponding galaxy absolute magnitude $M_{t,\rm UV}\sim-16$ in their case) and fit only $\beta$. We preserve the freedom of $M_t$ and use the priors to exclude extremely small $M_t$ to interpret the data.

The top panel of Fig. \ref{fig:2} shows the MCMC posterior distribution of the parameters. The bottom panel of Fig. \ref{fig:2} shows the results of the modification on SFE and the LF, respectively. The solid line represents scenario without turnover. The dashed line and the shaded region shows the best-fit results with our model and its 1 $\sigma$ uncertainties. The red dots with error bars represent the observational data from \cite{atek2018extreme}.

In Table \ref{tab:fit}, we list the $68$ \% intervals of three free parameters. $\beta$ is not well-constrained and the 1D posterior distribution is highly non-Gaussian. Therefore the 1D peak of the posterior and the multi-dimensional peak of the posterior can differ from the medians. Throughout this paper, we take the medians of the marginal distributions as our best-fits. The turnover mass $M_t$ is constrained at $2.17^{+2.42}_{-1.72}\times 10^{10} \rm M_{\odot}$. This translates into galaxies' absolute magnitude at $-16.46^{-1.14}_{+2.49}$. Note $M_t$ is where the SFE starts to decrease, different from where the LF peaks ($\frac{d\phi}{dM_h}|_{M_h=M_{h,\rm peak}}=0$) as \cite{atek2018extreme} and \cite{park2020turnover} adopted. In our results, the LF peaks at $M_{\rm UV,peak}\gtrsim -13.99-2.45$, correspondingly, $M_{h,\rm peak}\lesssim (4.57+20.03)\times 10^{9} \rm M_{\odot}$. This is consistent with $M_{\rm UV,peak}=-14.93^{+0.61}_{-0.52}$ \citep{atek2018extreme} while slightly higher than $log_{10}(M_{h,\rm peak})=9.40^{+0.18}_{-0.36}$ \citep{park2020turnover}. It is also consistent with the result that \cite{bouwens2022turnover} recently rule out a potential turnover brightward of $-14.3$ mag at $z\sim6$. The curvature parameter $\beta=2.17^{+2.42}_{-1.72}$, which suggests a strong turnover at the faint-end. However, both the constraints of the turnover mass and the curvature parameter are not particularly tight due to the limit of the data. Furthermore, the conversion factor $l_{\rm UV}=9.33^{+0.44}_{-0.42}\times 10^{27}~\rm erg~yr~s^{-1}M_{\odot}^{-1}$. This is in good agreement with the predictions derived from a stellar synthesis population model with a Salpeter initial stellar mass function in the range of $0.1$-$100 \rm M_{\odot}$, a constant star-formation rate and an evolving stellar metallicity \citep{madau2014cosmic,furlanetto2017minimalist,sabti2022galaxy}.

\begin{figure}
    \centering  
    \includegraphics[width=\columnwidth]{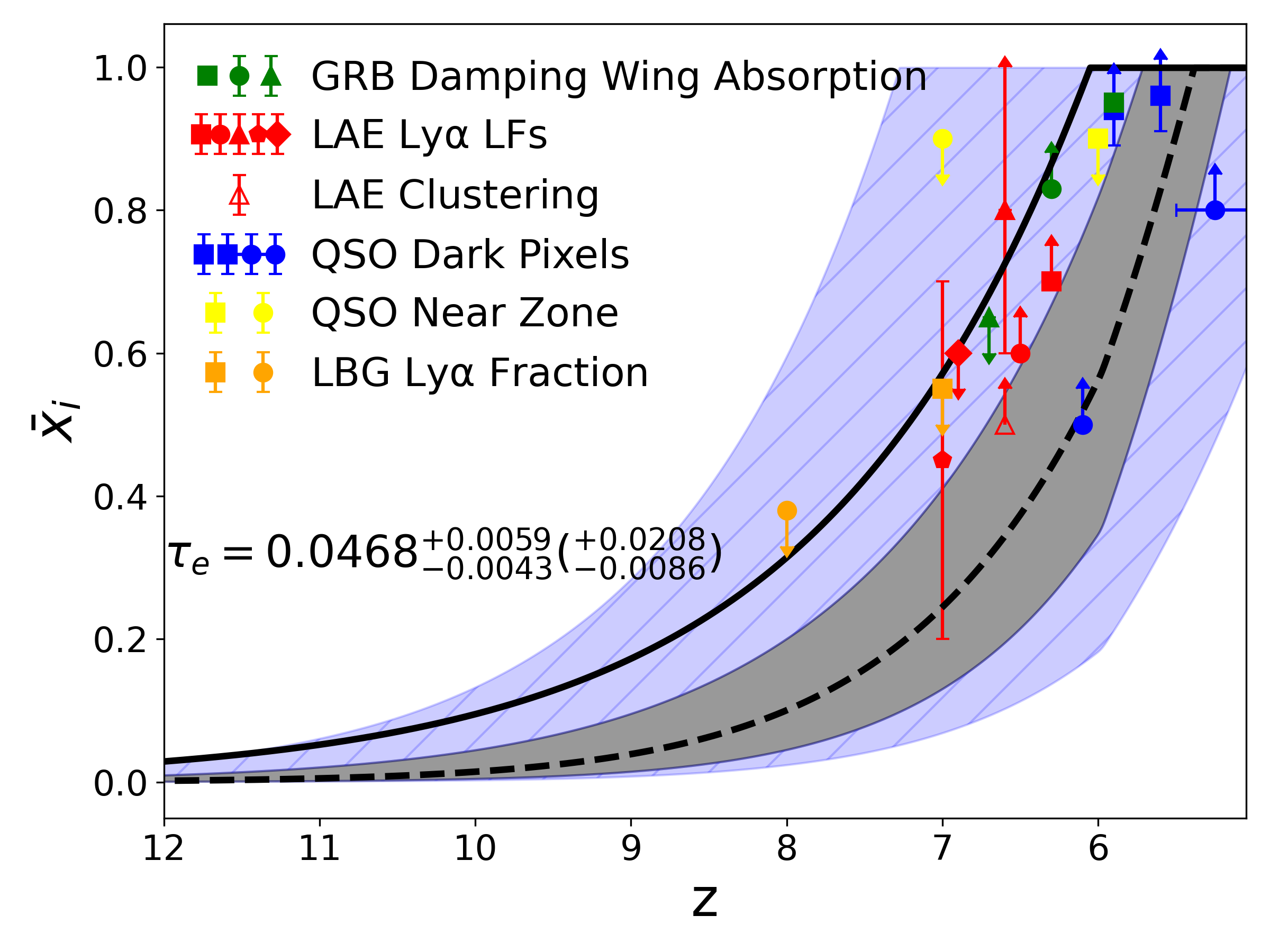}
    \caption{The evolution of the volume-averaged ionized hydrogen fraction. Again, the grey shaded area corresponds to the interval when we take the 68\% CL boundary of parameter sets. The dark solid line is our fiducial model and the black dashed line represents when we apply the best fit parameters. Blue shaded area further includes the uncertainty of $f_{esc,\rm LyC}$ estimated from \citet{meyer2020scape}. At the lower left is the CMB optical depth $\tau_e$ corresponding to the grey area (values in the bracket represent the blue area). Previous constraints on hydrogen neutral fraction using various probes are shown. The filled green square, circle and triangle indicate estimates based on GRB damping wing absorption by \citet{totani2016damping}, \citet{totani2006damping} and \citet{greiner2009damping}. The filled red square, circle, triangle, pentagon and diamond show the constraints from the LAE $\rm Ly\alpha$ LFs by \citet{malhotra2004lae}, \citet{kashikawa2011lae}, \citet{ouchi2010lae}, \citet{konno2014lae} and \citet{ota2017lae}. The open red triangle is from the clustering of LAEs \citep{ouchi2010lae}. The blue squares and circles denote the results from QSO dark $\rm Ly\alpha$ forest pixels by \citet{mcgreer2015darkfraction} and \citet{mcgreer2011dark}. The yellow square and circle are from sizes of QSO near zones by \citet{schroeder2013nearzone} and \citet{bolton2011nearzone}. The orange square and circle show results using the fraction of Ly$\alpha$ emitting LBGs at $z\sim7$ (joint constraints from \citealt{pentericci2014lbg,schenker2014lbg,caruana2014lbg,furusawa2016lbg}) and at $z\sim8$ from \citet{schenker2014lbg}.}
\label{fig:ionization}
\end{figure}

\subsection{Global 21-cm signal}

Fig.~\ref{fig:3} presents the predictions on the global 21-cm signal. From this figure, we can find that our modification on SFE causes later features on the global 21-cm signal with a deeper trough ($-230~\rm mK\gtrsim \delta T_b\gtrsim-250~\rm mK$) occurring at higher frequencies ($110~\rm MHz\lesssim\nu\lesssim150~\rm MHz$). The trough locates at $\sim$ $134^{+10}_{-17}$ $\rm MHz$ and the amplitude is $\sim$ $-237^{-6}_{+7}$ $\rm mK$ according to parameter 68 \% intervals, compared to (106\rm MHz, -212\rm mK) in our fiducial model.

We fit our global signal to the flattened Gaussian profile \citep{bowman2018absorption}, 
\begin{equation}
    \delta T_b (\nu)=-A\left (\frac{1-e^{-\tau e^B}}{1-e^{-\tau}} \right ), 
\label{eq:gauss}
\end{equation}
where,
\begin{equation}
    B=\frac{4(\nu-\nu_0)^2}{w^2} \ln\left [-\ln \left (\frac{1+e^{-\tau}}{2}\right )/\tau \right ],
\end{equation}
A is the absorption amplitude, $\nu_0$ is the centre frequency, $w$ is the full width at half maximum, and $\tau$ is a fattening factor. We fit these four parameters using MCMC method with flat priors. The best confidences are $(A, \nu_0, w, \tau)=(218.49^{+26.08}_{-23.76}\rm mK,129.02^{+3.74}_{-3.45}\rm MHz,81.16^{+9.98}_{-9.03},2.57^{+3.47}_{-1.76})$, shown as red dotted curve in Fig. \ref{fig:3}.

Because only the contributions of the known stellar population are included in our model and the results are extrapolated to higher redshifts, our predictions can be biased. The peak and amplitude in absorption is different from the measurement of the EDGES with $(A, \nu_0, \tau)=(530~\rm mK, 78.1~\rm MHz, 7)$ \citep{bowman2018absorption}. Recent results by SARAS \citep{singh2021detection} claim that the best-fit profile of EDGES is rejected with $95.3\%$ confidence in the $55-85~\rm MHz$ band, suggesting that further experiments are needed for the measurement of the global 21-cm signal. Moreover, comparing with the peak in absorption of $(A, \nu_0)=(160~\rm mK, 110~\rm MHz)$ by \cite{mirocha2017global}, the turnover in LF of the current stellar population models can delay the end of EoR, making deeper absorption.

Most models predict a relatively weak emission bump during EoR. This feature is a result of decreasing neutral fraction and rising spin temperature at the end of EoR. However, there are only weak emission features in our model. The reionization process ends fast in a range of $\Delta z \sim 2$ during which the IGM has not been properly heated to surpass the temperature of cosmic background radiation $T_{\gamma}$. This is because that the radiation background of UV and X ray sources is naturally underestimated as we conservatively take known galaxies into consideration, instead of the whole baryonic collapsed fraction \citep{mirocha2014decoding,gu2020direct}. The latter acts as a upper limit of galactic emissivity $\epsilon_{\nu}$ while the former acts as a lower limit. Smaller population of high-z faint galaxies will further deepen this limit. The major sources of the LyC photons responsible for reionization are yet not fully identified. Recently \citet{jiang2020quasar} claim that the quasar population can only provide less than 7 \% of the total photons and the star-forming galaxies play the major role.

In Fig. \ref{fig:ionization}, we present the evolution of $\Bar{x}_i$. The reionization ends at $z\sim5.4_{-0.2}^{+0.3}$ comparing to $\sim6.1$ in the absence of turnover. The corresponding CMB optical depth $\tau_e=0.0468^{+0.0059}_{-0.0043}$. When including the uncertainty of $f_{\rm esc,\rm LyC}=0.23^{+0.46}_{-0.11}$ \citep{meyer2020scape}, $\tau_e=0.0468^{+0.0208}_{-0.0086}$. This is consistent with other observations, for example, $z\simeq6$ from the Gunn-Peterson trough \citep{fan2006,bouwens2015reionization} and $\tau_e=0.0540^{+0.0074}_{-0.0074}$ from \cite{collaboration2020planck}. Our results are also in agreement with current constraints on neutral fraction at several redshifts, e.g., studies from quasar (QSO) dark $\rm Ly\alpha$ forest pixels at $z\sim5-6$ \citep{mcgreer2011dark,mcgreer2015darkfraction}, the LAE (lyman-alpha emitter) $\rm Ly\alpha$ LFs at $z\sim6-7$ \citep{malhotra2004lae,kashikawa2011lae,ouchi2010lae,konno2014lae,ota2017lae}, the clustering of LAEs at $z\sim6$ \citep{ouchi2010lae}, GRB (gamma-ray burst) damping wing absorption at $z\sim6-7$ \citep{totani2006damping,totani2016damping,greiner2009damping}, QSO near zones at $z\sim6,7$ \citep{schroeder2013nearzone,bolton2011nearzone} and $\rm Ly\alpha$ emitting LBGs (lyman-break galaxy) at $z\sim7,8$ \citep{pentericci2014lbg,schenker2014lbg,caruana2014lbg,furusawa2016lbg}. Although slight tension is noticeable, it can be solved by taking other ionizing sources into consideration or boosting the escape fraction of LyC photons.

\begin{figure*}
    \centering
    \includegraphics[width=0.47\textwidth]{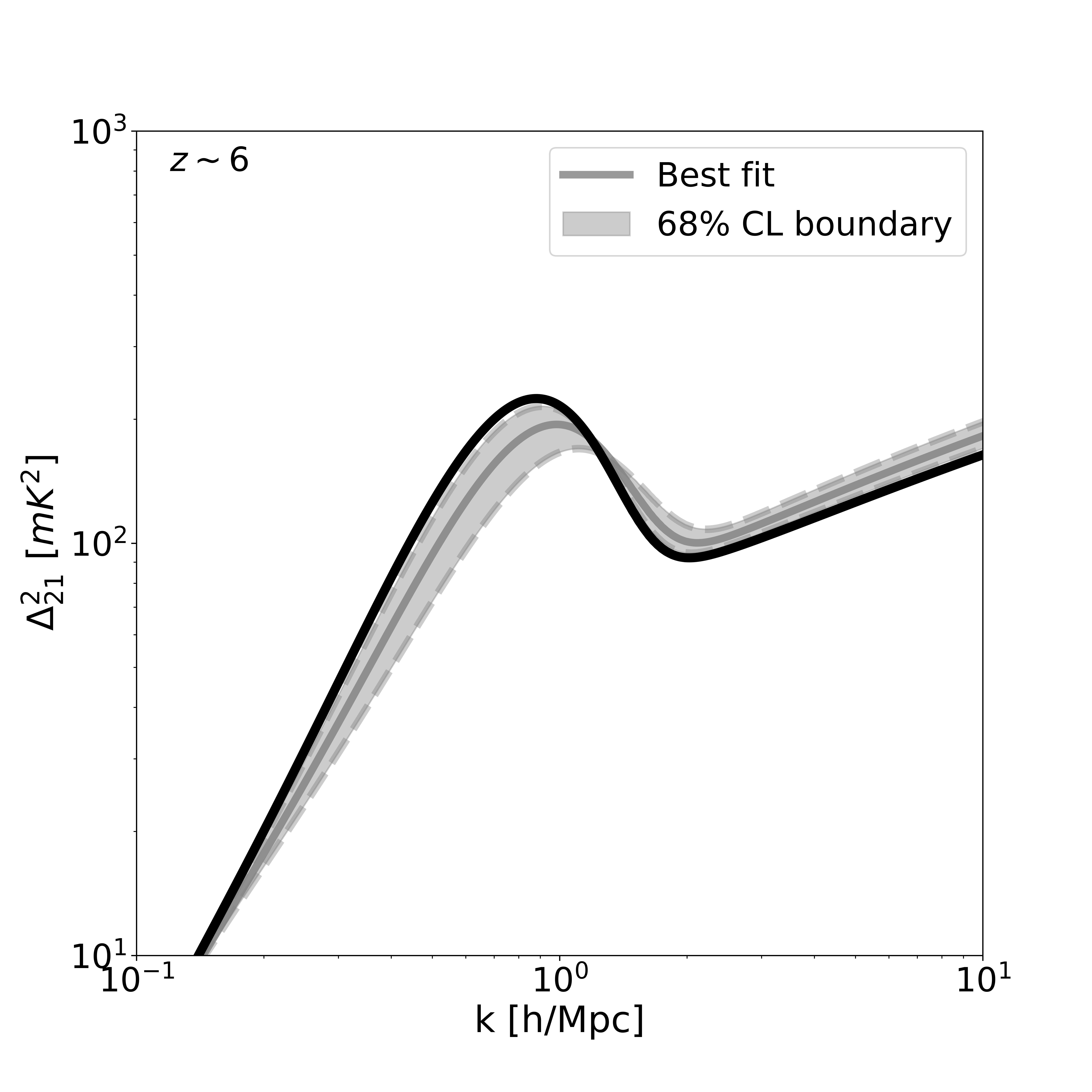}
    \includegraphics[width=0.47\textwidth]{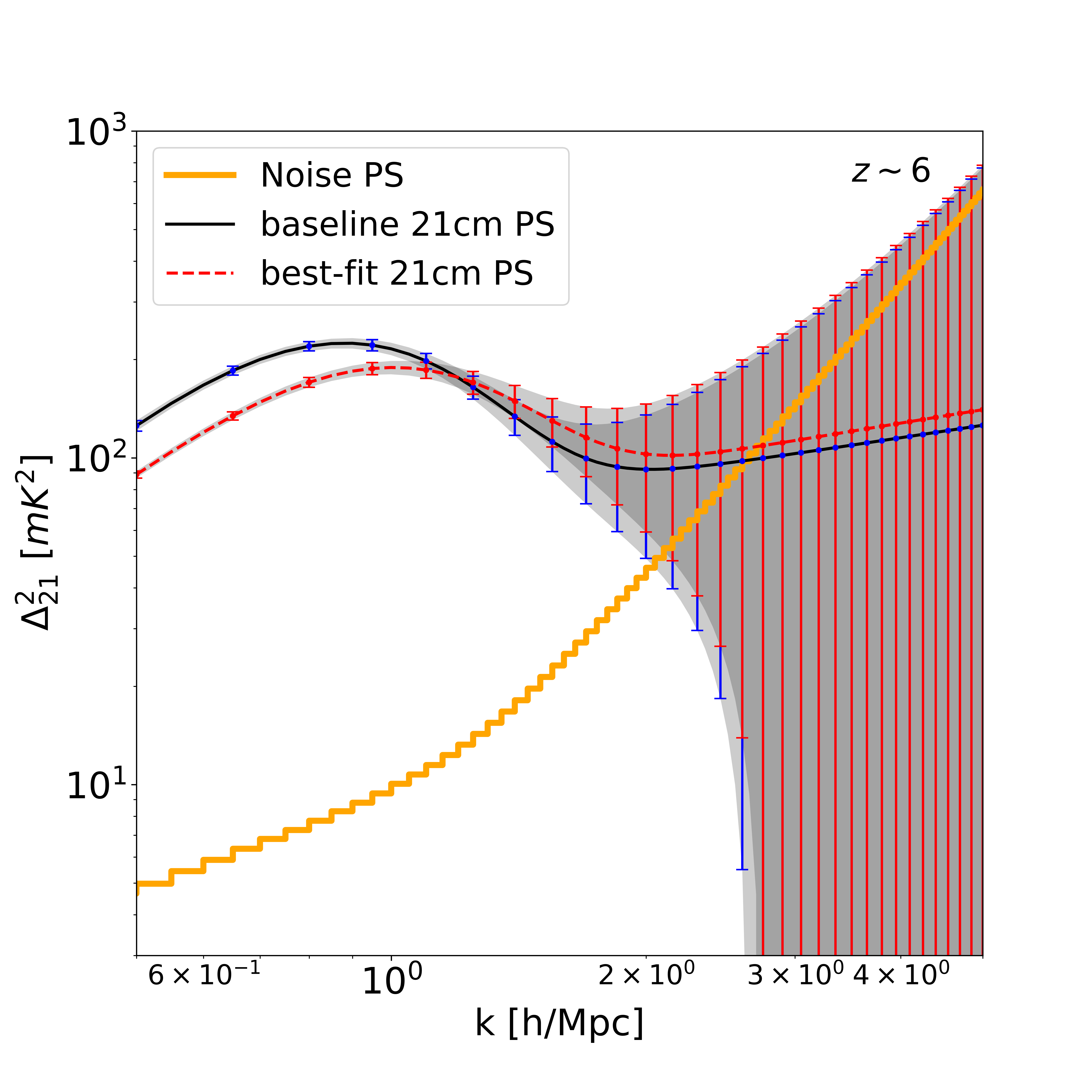}
    \caption{The 21-cm power spectra at z $\sim$ $6$. Left: The solid dark curve represents our baseline scenario. Grey curves apply our fitting: best fit (solid) and the 68\% CL boundary (dashed). Right: Predicted constraints on our EoR model assuming a 1000 hr observation with SKA1-low. The error bars represent 1$\sigma$ noise.}
    \label{fig:4}
\end{figure*}

\subsection{21-cm power spectrum}

In this section, we exam the impact of LF turnover on the 21-cm power spectrum as shown in the left panel of Fig.~\ref{fig:4}, by converting the turnover of LF into HMF. The thick line denotes the ST HMF, while in the shaded region, we multiply its low mass end according to Eq. \ref{con:turnover}. From this figure, we can find that at $k\sim 1 h \rm Mpc^{-1}$, where the ionized bubbles imprint on the power spectrum, suppression of star formation dampens the power from the $\rm H_{II}$ region. This can be explained that the ionizing process in the IGM is dampened as we manually decrease the ionizing efficiency of faint sources. Therefore, similar to the delayed absorption in the global 21-cm signal in Fig. \ref{fig:3}, at $z$ $\sim$ 6 the amplitude of $\Delta^2_{21}$ is slightly suppressed near the characteristic bubble size and the peak moves to smaller scales. Therefore, this will increase the error constraining the shape of 21-cm power spectra.

We examine the prospects for detecting the 21-cm power spectrum with SKA. We use the publicly available {\it Python} module {\it 21cmSense \footnote{\url{https://github.com/jpober/21cmSense}}} \citep{pober2013,pober2014} to examine the expected instrumental noise. Given the configuration design of any array, {\it 21cmSense} derives the gridded $uv$-visibilities to yield the noise power spectrum. For SKA1-low, it consists of 512 40m antennae stations. Current design plan calls for $\sim 40 \%$ of the field stations of a size $\rm d=40~m$ to lie in a compact core of $D\sim1~\rm km$. The noise in every $uv$ bin is calculated as,
\begin{equation}
    \Delta^2_{uv}(k) = X^2Y\frac{k^3}{2\pi^2}\frac{\Omega}{2t_0}T^2_{sys}
\end{equation}
where $X^2Y$ is a scalar translating observed units to cosmological distances in $h^{-1}$ Mpc, $\Omega$ is the solid angle of the primary beam of one element, and $t_0$ is the total amount of observation time which we set as 1000 hr. $T_{sys}$ is the total system temperature modeled as, \citep{deboer2017hydrogen},
\begin{equation}
    T_{sys}(\nu) = 100~\rm K+120~\rm K\times(\frac{\nu}{150\rm MHz})^{-2.55}
\end{equation}

Combined with thermal noise, the cosmic variance contribution to the error is estimated from the signal $\Delta^2_{21}$. Assuming Gaussian distribution for the variance, the total noise power is then an inversely weighted summation over all the individual modes,
\begin{equation}
    \Delta^2_N(k) = \left [\sum_i \frac{1}{(\Delta^2_{uv}(k)+\Delta^2_{21}(k))^2} \right ]^{-1/2}
\end{equation}

Foreground mitigation is performed via avoidance, where the 21-cm signal is restricted outside of the contaminated foreground wedge. In 2D cylindrical space, the wedge is defined as,
\begin{equation}
    k_{\parallel} = ak_{\perp} + b
\end{equation}
where $k_{\parallel}$ and $k_{\perp}$ are the line-of-sight and transverse modes in Fourier space. In this work, we consider only the moderate foreground scenario treating the buffer $b=0.1h \rm Mpc^{-1}$.

In the right panel of Fig.~\ref{fig:4}, we show the power spectrum constraints with $1\sigma$ noise error. We find that comparing to our baseline model, the best fitting power spectrum suffers from slightly lower SNR at the peak while on small and large scales the errors are nearly the same. The results infer that it is possible to distinguish these two scenarios assuming basic foreground contamination with SKA.

\section{Conclusions}
\label{Conclusions}

The HFF has delivered the deepest observation of lensing clusters to date and push the limit down to unprecedentedly $M_{\rm UV} \sim-13$ at $z\sim6$. The most interesting finding is the small population of high-redshift faint galaxies. This may reveal a different picture of the early universe and the first stars. Based on the past literature \citep{mirocha2017global}, our work further investigates the impact of the turnover from the current galaxy population models on the reionization history. We use the observations from the HFFs to constrain the UV LF model and summarize our main findings as follows:

(i) The observed very-faint-end turnover feature of high-z galaxy UV LF is not expected according to conventional wisdom of galaxy luminosity. However, this feature can be successfully interpreted by inefficiency of star formation on small halos. According to our model, the star formation is strongly suppressed in halos smaller than $M_t=2.17^{+2.42}_{-1.72}\times10^{10} \rm M_{\odot}$ for $68$ \% CL, corresponding to galaxies faintwards of $M_{\rm UV}$ $\sim$ $-16.46^{-1.14}_{+2.49}$. The UV LF peaks at $\rm \sim -13.99$, which translates into halo mass $M_{h,\rm peak}\sim 4.57\times 10^{9} \rm M_{\odot}$.

(ii) We find that, in our reionization paradigm, the time point of X-ray heating will significantly delay. Besides, the IGM prior to the EoR would be slightly colder than the baseline model, as a result of the lack of heating sources. The absorption trough in the global 21-cm signal is sensitive to both of our SFE model parameters, typically with a $\sim$ $30$ MHz frequency spread and a $\sim$ $-15$ mK amplitude spread given our $68$ \%  fit between a lower limit where $\beta=4.59,M_t=4.68\times10^{10} \rm M_{\odot}$ and an upper limit where $\beta=0.45,M_t=7.4\times10^{9} \rm M_{\odot}$.

(iii) Furthermore, we considered the impact of UV LF on the 21-cm power spectrum. We find that the 21-cm field fluctuation at $z$ $\sim6$ is sensitive to star formation in faint galaxies. According to our best fit, the peak amplitude moves down from $\sim224$ $\rm mK^2$ down to $\sim215$ $\rm mK^2$ and shifts towards smaller scales from 0.86 $h \rm Mpc^{-1}$ to 0.91 $h \rm Mpc^{-1}$. Such impact is able to be detected with SKA assuming a 1000 hr observation and a moderate foreground.

The results of this paper demonstrate the impact of the turnover in LF on the EoR 21-cm signal. The upcoming optical/infrared experiments, including James Webb Space Telescope (JWST), Roman Space Telescope (RST; \citealt{spergel2015wide}) and the China Space Station Telescope (CSST) Multi-channel Imager (M; \citealt{cao2022calibrating}), will have the capability to provide tighter constraints on the faint-end LF. This can allow the implementation on the direct detection of the CD/EoR via the redshifted HI 21-cm signal by the low frequency experiments (e.g., MWA, LOFAR, HERA, SKA).

\section*{Acknowledgements}

We acknowledge support from the Ministry of Science and Technology of China (grant Nos. 2020SKA0110100), the National Science Foundation of China (11973069, 11973070), the Shanghai Committee of Science and Technology grant No.19ZR1466600 and Key Research Program of Frontier Sciences, CAS, Grant No. ZDBS-LY-7013. HYS, QZ and QG thank the Supports of Shanghai International Science projects: the SKA Science Regional Centres international partners project (No. 19590780200).

\section*{Data availability}

All the code and data used to produce the analysis in this article are publicly available.




\bibliographystyle{mnras}
\bibliography{ref} 




\appendix


\bsp	
\label{lastpage}
\end{document}